\documentclass [twoside,12pt] {article}
\usepackage{setspace,graphicx,natbib,bm,fancyhdr}
\usepackage{amsmath,amsthm,amssymb,latexsym}
\usepackage{color}
\usepackage{graphicx}

\usepackage{authblk}

\newcommand{\rf}{\vskip .1in\par\sloppy\hangindent=1pc\hangafter=1
                 \noindent}

\newcommand{\ch}[1]{\mbox{$\stackrel{\sim}{#1}$}}
\newcommand{\vct}[1]{\mbox{$\stackrel{\rightharpoonup}{#1}$}}
\newcommand{\tld}[1]{\mbox{$\stackrel{\sim}{#1}$}}

\newcommand{\slas}[1]{\mbox{${{#1} \!\!\! /}$}}

\begin{document}
\title{\bf Divergence free quantum field theory using a spectral calculus of Lorentz invariant measures} 
\author{John Mashford \\
School of Mathematics and Statistics \\
University of Melbourne, Victoria 3010, Australia \\
E-mail: mashford@unimelb.edu.au}
\date{\today}
\maketitle

\begin{abstract}
 
This paper presents a spectral calculus for computing the spectrum of a causal Lorentz invariant Borel complex measure on Minkowski space, thereby enabling one to compute the density for such a measure with respect to Lebesque measure. It is proved that the convolution of arbitrary causal Lorentz invariant Borel measures exists and the product of such measures exists in a wide class of cases. Techniques for their computation are presented. Divergent integrals in quantum field theory (QFT) are shown to have a well defined existence as Lorentz covariant measures. The case of vacuum polarization is considered and the spectral vacuum polarization function is shown to have very close agreement with the vacuum polarization function obtained using dimensional regularization / renormalization in the timelike domain. Using the spectral vacuum polarization function the exact Uehling potential function is derived. The spectral running coupling constant is computed and is shown to converge for all energies while the integral defining the running coupling constant obtained using dimensional regularization / renormalization is shown to diverge for all non-zero energies. 

\end{abstract}

\tableofcontents

\listoffigures

\section{Introduction}

The divergences in quantum field theory (QFT) are currently generally dealt with by using the techniques of regularization and renormalization.  Two of the principal methods of regularization are Pauli-Villars regularization and dimensional regularization. Both of these methods involve modifying a divergent integral to form an integral which exists in a manner depending on a parameter where the parameter is the mass $\Lambda$ of a fictitious photon for Pauli-Villars regularization or a perturbation $\epsilon>0$ of the space-time dimension $D=4-\epsilon$ for dimensional regularization. The parameter is then varied towards the value that it would have if the divergent integral existed ($\Lambda\rightarrow\infty$ for Pauli-Villars, $\epsilon\rightarrow0$ for dimensional) and the badly behaved contributions (e.g. terms of the order of $\log(\Lambda)$ or $\epsilon^{-1}$) are subtracted out or ignored to obtain finite answers which can be compared with experiment. The process of removing the badly behaved contributions is called renormalization (e.g. minimal subtraction).

Many of the initial developers of QFT such as Dirac and Feynman were not happy with the fact that many of the integrals in QFT, in particular, those involving fermion loops, do not exist in the mathematical sense but more recently, especially through the important work of Wilson and others on the renormalization group, the parameterized variation with scale has been seen to have physical significance and not just the result of a mathematical artifice. In particular the concept of running coupling constant and the comparison of its behavior in QED and QCD, which is asymptotically free, is seen to be of great physical significance. The significance has propagated into other areas of physics such as solid state physics and critical phenomena in condensed matter physics.

We believe, nevertheless, that it would be desirable to have well defined initial equations or principles as a starting point for physical theory which are such that concepts such as the running coupling constant would follow from these basic principles. We have shown in a previous paper (Mashford, 2017b) how one can, through a brief formal argument, consider the problematic objects in QFT as being Lorentz covariant Borel complex measures (more generally $K$ covariant ${\bf C}^{4\times4}$ valued measures) on Minkowski space. We will repeat this derivation in the present paper for the case of the contraction of the vacuum polarization tensor. Having given a definition of the objects as well defined mathematical objects one can proceed and analyze these objects, computing the consequences of assuming them, without infinities or ill-definedness propagating through the calculations.

It can be shown (see Appendix 2) that any Lorentz invariant Borel complex measure on Minkowski space has a certain spectral representation.
An important part of this paper is the presentation of a spectral calculus in Section~\ref{section:spec_calculus} whereby the spectrum of a causal Lorentz invariant Borel measure on Minkowski space can be calculated, where by causal is meant that the support of the measure is contained in the closed future null cone of the origin.

If, using the spectral calculus, one can obtain a spectrum for a causal Lorentz invariant Borel measure which is a continuous function (or, more generally, a sufficiently well behaved measurable function) then, as we will show, one can compute an equivalent density for the measure with respect to Lebesque measure on ${\bf R}^4$ which can be used in QFT calculations. 

We will show, generally, how to convolve or form products of causal Lorentz invariant Borel measures using their spectral representations. This is to be compared to the work of Scharf and others, dating back to the paper of Epstein and Glaser, (1973)  on forming products of causal distributions.

The concept of spectral representation in QFT dates back to the work of K\"{a}llen (1952) and Lehmann (1954) who, independently, proposed the representation
\begin{equation}
<0|[\phi(x),\phi^{\dagger}(y)]|0>=i\int_0^{\infty}d{m^{\prime}}^2\sigma({m^{\prime}}^2)\Delta_{m^{\prime}}(x-y),
\end{equation}
for the commutator of interacting fields where $\Delta_{m^{\prime}}$ is the Feynman propagator corresponding to mass $m^{\prime}$. Itzykson and Zuber (1980) state, with respect to $\sigma$, ``In general this is a positive measure with $\delta$-function singularities." While K\"{a}llen, Lehmann and others propose and use this decomposition they do not present a way to compute the spectral measure $\sigma$. As mentioned above one of the main results of the present paper is a presentation of the spectral calculus which enables one to compute the spectral function of a causal Lorentz invariant Borel measure on Minkowski space. This spectral calculus is quite easy to use in practice but it is somewhat tedious to prove rigorously its validity (see Appendix 4).

In Section~\ref{section:prop_conv_spec} of the paper we use the spectral calculus and other methods to compute the spectrum of the measure $\Omega_m*\Omega_m$ which is a convolution of the standard Lorentz invariant measure on the mass $m$ mass shell (i.e the Feynman propagator corresponding to mass $m$ on the space of positive energy functions) with itself, where $m>0$. In Section~\ref{section:imaginary_prop_conv_spec} we use general arguments to compute the spectrum of $\Omega_{im}*\Omega_{im}$ where $\Omega_{im}$ is standard Lorentz invariant measure on the imaginary mass hyperboloid corresponding to mass $im$, $m>0$. In Section~\ref{section:density} we show how an equivalent density for a Lorentz invariant measure can be computed from its spectrum. In Section~\ref{section:tensor} we describe how the spectral theorem can be generalized to Lorentz covariant tensor valued measures. In Section~\ref{section:conv_and_products} we describe how the convolution and product of Lorentz invariant measures can be computed from their spectra. These computations form practice for the main application of the paper which is an investigation in Section~\ref{section:vacuum_polarization} of vacuum polarization, i.e. the self energy of the photon.

In Section~\ref{section:vacuum_polarization} we compute the spectral function and hence the density associated with the complex measure obtained by contracting the vacuum polarization tensor. This is used to define our spectral vacuum polarization function. Our function is seen to agree with a high degree of accuracy (up to finite normalization) with the vacuum polarization function obtained using regularization/renormalization. 

In Section~\ref{section:Uehling} we follow Weinberg and others' method for the computation of the Uehling contribution to the Lamb shift in the H atom. Ours differs because we have a different vacuum polarization function in the imaginary mass (spacelike) regime. We derive, using the Born approximation, the Uehling potential function from which the Uehling effect for the hydrogen atom can be exactly computed.

In Section~\ref{section:running_coupling} we compute and display the running coupling constant for 1 loop QED. This computation is shown to be convergent when the spectral vacuum polarization function is used while the computation using the vacuum polarization function obtained using dimensional regularization / renormalization is shown to be divergent for all non-zero energies.

The paper concludes in Section~\ref{section:conclusion}. 

\section{A spectral calculus of Lorentz invariant measures\label{section:spec_calculus}}

Let ${\mathcal B}_0({\bf R}^4)=\{\Gamma\in{\mathcal B}({\bf R}^4):\mbox{ $\Gamma$ is relatively compact}\}$ where ${\mathcal B}({\bf R}^4)$ is the Borel algebra of ${\bf R}^4$. By a Borel complex measure on Minkowski space we will mean a map $\mu:{\mathcal B}_0({\bf R}^4)\rightarrow{\bf C}$, where ${\bf C}$ is the complex numbers, such that for all $C\in{\mathcal B}_0({\bf R}^4)$ the map $\Gamma\mapsto\mu(\Gamma)$ for $\Gamma\in{\mathcal B}(C)$ is a Borel complex measure on $C$ in the usual sense of the term (Halmos, 1988). For the rest of this paper the term `measure' will mean `Borel measure'.

A measure $\mu:{\mathcal B}_0({\bf R}^4)\rightarrow{\bf C}$ is Lorentz invariant if
\begin{equation}
\mu(\Lambda\Gamma)=\mu(\Gamma),\forall\Lambda\in O(1,3)^{\uparrow+},\Gamma\in{\mathcal B}_0({\bf R}^4).
\end{equation}

Consider the following general form of a complex measure $\mu$ on Minkowski space.
\begin{equation} \label{eq:invariant1}
\mu(\Gamma)=c\delta(\Gamma)+\int_{m=0}^{\infty}\Omega^{+}_m(\Gamma)\,\sigma_1(dm)+\int_{m=0}^{\infty}\Omega_m^{-}(\Gamma)\,\sigma_2(dm)+\int_{m=0}^{\infty}\Omega_{im}(\Gamma)\,\sigma_3(dm),
\end{equation}
where $c\in{\bf C}$ (the complex numbers), $\delta$ is the Dirac delta function (measure), $\sigma_1,\sigma_2,\sigma_3:{\mathcal B}_0([0,\infty))\rightarrow{\bf C}$ are complex measures, $\Omega_m^{+}$ is the standard Lorentz invariant measure concentrated on the mass shell $H_m^{+}$ (see (Mashford, 2017b)), $\Omega_m^{-}$ is the standard Lorentz invariant measure concentrated on the mass shell $H_m^{-}$ and $\Omega_{im}$ is the standard Lorentz invariant measure on the imaginary mass hyperboloid $H_{im}$. Then $\mu$ is a Lorentz invariant measure. Conversely we have the following.
\newtheorem{theorem}{Theorem}
\begin{theorem}{The Spectral Theorem.}
Let $\mu:{\mathcal B}_0({\bf R}^4)\rightarrow{\bf C}$ be a Lorentz invariant complex measure. Then $\mu$ has the form of Eq.~\ref{eq:invariant1} for some $c\in{\bf C}$ and  spectral measures $\sigma_1,\sigma_2$ and $\sigma_3$.
\end{theorem}
A proof of this theorem is given in Appendix 2.

If $\sigma_2=\sigma_3=0$ then $\mu$ will be said to be {\em causal} or a type I measure. If $\sigma_1=\sigma_3=0$ then $\mu$ will be said to be a type II measure and if $c=0$ and $\sigma_1=\sigma_2=0$ then $\mu$ will be said to be a type III measure. Thus any Lorentz invariant measure is a sum of a type I measure, a type II measure and a type III measure. In particular, any measure of the form
\begin{equation} \label{eq:invariant2}
\mu(\Gamma)=\int_{m=0}^{\infty}\sigma(m)\Omega^{+}_m(\Gamma)\,dm,
\end{equation} 
where $\sigma:(0,\infty)\rightarrow{\bf C}$ is a locally integrable function and the integration is carried out with respect to the Lebesgue measure, is a causal Lorentz invariant complex measure. If $\sigma$ is polynomially bounded then $\mu$ is a tempered measure.

The spectral calculus that we will now explain is a very simple way to compute the spectrum $\sigma$ of a Lorentz invariant measure $\mu$ if we know that $\mu$ can be written in the form of Eq.~\ref{eq:invariant2} and $\sigma$ is continuous.

For $m>0$ and $\epsilon>0$ let $S(m,\epsilon)$ be the hyperbolic (hyper-)disc defined by
\begin{equation}
S(m,\epsilon)=\{p\in{\bf R}^4:p^2=m^2,|{\vct p}|<\epsilon, p^0>0\},
\end{equation}
where, as usual in QFT, $p^2=\eta_{\mu\nu}p^{\mu}p^{\nu}=(p^0)^2-(p^1)^2-(p^2)^2-(p^3)^2$ and ${\vct p}=\pi(p)=(p^1,p^2,p^3)$. For $a,b\in{\bf R}$ with $0<a<b$ let $\Gamma(a,b,\epsilon)$ be the hyperbolic cylinder defined by
\begin{equation}
\Gamma(a,b,\epsilon)=\bigcup_{m\in(a,b)}S(m,\epsilon).
\end{equation}
Now suppose that we have a measure in the form of Eq.~\ref{eq:invariant2} where $\sigma$ is continuous. Then we can write (using the notation of (Mashford, 2017b))
\begin{eqnarray}
\mu(\Gamma(a,b,\epsilon)) & = & \int_{m=0}^{\infty}\sigma(m)\Omega_m(\Gamma(a,b,\epsilon))\,dm \nonumber \\
    & = & \int_{m=0}^{\infty}\sigma(m)\int_{\pi(\Gamma(a,b,\epsilon)\cap H_m^{+})}\frac{d{\vct p}}{\omega_m({\vct p})}\,dm \nonumber \\
    & = & \int_a^b\sigma(m)\int_{B_{\epsilon}({\vct 0})}\frac{d{\vct p}}{\omega_m({\vct p})}\,dm \nonumber \\
    & \approx & \frac{4}{3}\pi\epsilon^3\int_a^b\frac{\sigma(m)}{m}\,dm.  
\end{eqnarray}
where $B_{\epsilon}({\vct 0})=\{{\vct p}\in{\bf R}^3:{|\vct p}|<\epsilon\}$.

The approximation $\approx$ in the last line comes about because the function ${\vct p}\mapsto\omega_m({\vct p})$ is not constant over $B_{\epsilon}({\vct 0})$.

Thus if we define 
\begin{equation} \label{eq:limit1}
g_a(b)=g(a,b)=\lim_{\epsilon\rightarrow0}\epsilon^{-3}\mu(\Gamma(a,b,\epsilon)),
\end{equation}
then we can retrieve $\sigma$ using the formula
\begin{equation} 
\sigma(b)=\frac{3}{4\pi}bg_a^{\prime}(b).
\end{equation}
Thus we have proved the following fundamental theorem of the spectral calculus of causal Lorentz invariant measures.
\begin{theorem} \label{theorem:fundamental}
Suppose that $\mu$ is a causal Lorentz invariant measure with continuous spectrum $\sigma$. Then $\sigma$ can be calculated from the formula
\begin{equation}
\sigma(b)=\frac{3}{4\pi}bg_a^{\prime}(b),
\end{equation}
where, for $a,b\in{\bf R}, 0<a<b, g_a:(a,\infty)\rightarrow{\bf R}$ is given by Eq.~\ref{eq:limit1}.
\end{theorem} 
To make the proof of this theorem rigorous we prove the following.
\newtheorem{lemma}{Lemma}
\begin{lemma}
Let $a,b\in{\bf R}, 0<a<b$. Then
\begin{equation}
\lim_{\epsilon\rightarrow0}\epsilon^{-3}\int_{B_{\epsilon}(0)}\frac{d{\vct p}}{\omega_m({\vct p})}=\frac{4\pi}{3}\frac{1}{m},
\end{equation}
uniformly for $m\in[a,b]$.
\end{lemma}
{\bf Proof} 
Define
\begin{equation}
I=I(m,\epsilon)=\int_{B_{\epsilon}(0)}\frac{d{\vct p}}{\omega_m({\vct p})}.
\end{equation}
Then
\begin{equation}
I=\int_{r=0}^{\epsilon}\frac{4\pi r^2\,dr}{(r^2+m^2)^{\frac{1}{2}}}.
\end{equation}
Now
\[ I_1<I<I_2, \]
where
\[ I_1=\int_{r=0}^{\epsilon}\frac{4\pi r^2\,dr}{(\epsilon^2+m^2)^{\frac{1}{2}}}=\frac{4\pi}{(\epsilon^2+m^2)^{\frac{1}{2}}}\frac{1}{3}\epsilon^3, \]
\[ I_2=\int_{r=0}^{\epsilon}\frac{4\pi r^2\,dr}{m}=\frac{4\pi}{m}\frac{1}{3}\epsilon^3. \]
Therefore
\[ \frac{4\pi}{3(\epsilon^2+m^2)^{\frac{1}{2}}}<\epsilon^{-3}I<\frac{4\pi}{3m}. \]
Thus
\[ \frac{4\pi}{3m}-\frac{4\pi}{3(\epsilon^2+m^2)^{\frac{1}{2}}}>\frac{4\pi}{3m}-\epsilon^{-3}I>0. \]
Hence
\begin{equation}
\left|\epsilon^{-3}I-\frac{4\pi}{3m}\right|<\frac{4\pi}{3m}-\frac{4\pi}{3(\epsilon^2+m^2)^{\frac{1}{2}}}.
\end{equation}
We have
\begin{eqnarray}
\frac{4\pi}{3m}-\frac{4\pi}{3(\epsilon^2+m^2)^{\frac{1}{2}}} & = & \frac{4\pi}{3}\frac{(\epsilon^2+m^2)^{\frac{1}{2}}-m}{m(\epsilon^2+m^2)^{\frac{1}{2}}} \nonumber \\
    & = & \frac{4\pi}{3}\frac{\epsilon^2}{m(\epsilon^2+m^2)^{\frac{1}{2}}((\epsilon^2+m^2)^{\frac{1}{2}}+m)} \nonumber \\
    & < &  \frac{4\pi}{3}\frac{\epsilon^2}{2m^3} \nonumber \\
    & \leq & \frac{4\pi}{3}\frac{\epsilon^2}{2a^3}, \mbox{ for all }m\in[a,b]. \nonumber
\end{eqnarray}
Therefore
\begin{equation}
\left|\epsilon^{-3}I-\frac{4\pi}{3m}\right|<\frac{4\pi}{3}\frac{\epsilon^2}{2a^3},
\end{equation}
for all $m\in[a,b]$

 $\Box$

 This lemma justifies the step of taking the limit under the integral sign (indicated by the symbol  $\approx$) in the proof of Theorem~\ref{theorem:fundamental}.

More generally, suppose that $\mu:{\mathcal B}_0({\bf R}^4)\rightarrow{\bf C}$ is a causal Lorentz invariant measure on Minkowski space with spectrum $\sigma$. Then, by the Lebesgue decomposition theorem there exist unique measures $\sigma_c,\sigma_s:{\mathcal B}_0([0,\infty))\rightarrow{\bf C}$ such that $\sigma=\sigma_c+\sigma_s$ where $\sigma_c$, the continuous part of the spectrum of $\mu$, is absolutely continuous with respect to Lebesque measure and $\sigma_s$, the singular part of the spectrum of $\mu$, is singular with respect to $\sigma_c$.  

It is straightforward to prove the following.
\begin{theorem} \label{Theorem:Th2}
Suppose that $a^{\prime},b^{\prime}\in{\bf R}$ are such that $0<a^{\prime}<b^{\prime}$, $\sigma_c|_{(a^{\prime},b^{\prime})}$ is continuous. Then for all $a,b\in{\bf R}$ with $a^{\prime}<a<b<b^{\prime}$, $g_a(b)$ defined by Eq.~\ref{eq:limit1} exists and is continuously differentiable. Furthermore $\sigma_c|_{(a^{\prime},b^{\prime})}$ can be computed using the formula
\begin{equation} \label{eq:retrieve_sigma_c}
\sigma_c(b)=\frac{3}{4\pi}bg_a^{\prime}(b),
\end{equation}
and
\begin{equation}
\sigma_s(E)=0, \forall \mbox{ Borel }E\subset(a^{\prime},b^{\prime}).
\end{equation}
Conversely suppose that $a^{\prime},b^{\prime}\in{\bf R}$ are such that $0<a^{\prime}<b^{\prime}$ and for all $a,b\in{\bf R}$ with $a^{\prime}<a<b<b^{\prime}$, $g_a(b)$ defined by Eq.~\ref{eq:limit1} exists and is continuously differentiable. Then $\sigma_c|_{(a^{\prime},b^{\prime})}$ is continuous and can be retrieved using the formula of Eq.~\ref{eq:retrieve_sigma_c}.
\end{theorem}

\section{Investigation of the measure defined by the convolution $\Omega_m*\Omega_m$\label{section:prop_conv_spec}}

\subsection{Determination of some properties of $\Omega_m*\Omega_m$}

Consider the measure defined by
\begin{equation}
\mu(\Gamma)=(\Omega_m*\Omega_m)(\Gamma)=\int\chi_{\Gamma}(p+q)\,\Omega_m(dp)\,\Omega_m(dq),
\end{equation}
where, for any set $\Gamma$, $\chi_{\Gamma}$ denotes the characteristic function of $\Gamma$ defined by
\begin{equation}
\chi_{\Gamma}(p)=\left\{\begin{array}{l}
1\mbox{ if }p\in\Gamma \\
0\mbox{ otherwise.}\end{array}\right.
\end{equation}
$\mu$ exists as a measure because as $|p|,|q|\rightarrow{\infty}$ with $p,q\in H_m^{+}$, $(p+q)^0\rightarrow{\infty}$ and so $p+q\mbox{ is eventually }{\slas \in}\Gamma$ for any compact set $\Gamma\subset{\bf R}^4$. Now
\begin{eqnarray}
\mu(\Lambda(\Gamma)) & = & \int\chi_{\Lambda(\Gamma)}(p+q)\,\Omega_m(dp)\,\Omega_m(dq) \nonumber \\
    & = & \int\chi_{\Gamma}(\Lambda^{-1}p+\Lambda^{-1}q)\,\Omega_m(dp)\,\Omega_m(dq) \nonumber \\
    & = & \int\chi_{\Gamma}(p+q)\,\Omega_m(dp)\,\Omega_m(dq) \nonumber \\
    & = & \mu(\Gamma),
\end{eqnarray}
for all $\Lambda\in O(1,3)^{+\uparrow}, \Gamma\in{\mathcal B}_0({\bf R}^4)$. Thus $\mu$ is a Lorentz invariant measure.

We will now show that $\mu$ is concentrated in the set 
\begin{equation}
C_{2m}=\{p\in{\bf R}^4:p^2\ge4m^2,p^0>0\},
\end{equation}
and therefore, that $\mu$ is causal.
Let $U\subset{\bf R}^4$ be open. Then
\begin{equation}
\mu(U)=\int_{{\bf R}^3}\int_{{\bf R}^3}\chi_U(\omega_m({\vct p})+\omega_m({\vct q}),{\vct p}+{\vct q})\,\frac{d{\vct p}}{\omega_m({\vct p})}\,\frac{d{\vct q}}{\omega_m({\vct q})}.
\end{equation} 
Therefore, using continuity, it follows that
\begin{eqnarray}
\mu(U)>0 & \Leftrightarrow & (\exists p\in U,{\vct q}_1,{\vct q}_2\in{\bf R}^3)\mbox{ } p=(\omega_m({\vct q}_1)+\omega_m({\vct q}_2),{\vct q}_1+{\vct q}_2). \nonumber
\end{eqnarray}
Suppose that $p\in\mbox{supp}(\mu)$ (the support of the measure $\mu$) i.e $p$ is such that $\mu(U)>0$ for all open neighborhoods $U$ of $p$. Let $U$ be an open neighborhood of $p$. Then, since $\mu(U)>0$, there exists $q\in U, {\vct q}_1,{\vct q}_2\in{\bf R}^3$ such that $q=(\omega_m({\vct q}_1)+\omega_m({\vct q}_2),{\vct q}_1+{\vct q}_2)$.  Clearly $q^0\ge2m$. Since this is true for all neighborhoods $U$ of $p$  it follows that $p^0\ge2m$. By Lorentz invariance we may assume without loss of generality that ${\vct p}=0$. Therefore $p^2\ge 4m^2$. Thus supp$(\mu)\subset C_{2m}$. 

For the converse, let $p=(\omega_m({\vct p}),{\vct p}), q= (\omega_m({\vct p}),-{\vct p})\in H_m^{+}$ for ${\vct p}\in{\bf R}^3$. As ${\vct p}$ ranges over ${\bf R}^3$, $p+q=(2\omega_m({\vct p}),{\vct 0})$ ranges over $\{(m^{\prime},{\vct 0}):m^{\prime}\geq2m\}$. It follows using Lorentz invariance that supp$(\mu)\supset C_{2m}$.

Therefore the support supp$(\mu)$ of $\mu$ is $C_{2m}$. Therefore by the spectral theorem $\mu$ has a spectral representation of the form
\begin{equation} \label{eq:convolution1}
\mu(\Gamma)=\int_{m^{\prime}=2m}^{\infty}\Omega_{m^{\prime}}(\Gamma)\,\sigma(dm^{\prime}),
\end{equation}
for some measure $\sigma:{\mathcal B}_0([2m,\infty))\rightarrow{\bf C}$. 

\subsection{Computation of the spectrum of $\Omega_m*\Omega_m$ using the spectral calculus}

Let 
\begin{equation}
\mu=\Omega_m*\Omega_m.
\end{equation}
For $a,b\in{\bf R}$ with $0<a<b$ and $\epsilon>0$ let 
\begin{equation}
g(a,b,\epsilon)=\mu(\Gamma(a,b,\epsilon)).
\end{equation}
We would like to calculate
\begin{equation}
g_a(b)=\lim_{\epsilon\rightarrow0}\epsilon^{-3}g_a(b,\epsilon),
\end{equation}
where
\begin{equation}
g_a(b,\epsilon)=g(a,b,\epsilon),
\end{equation}
and then retrieve the spectral function as
\begin{equation}
\sigma(b)=\frac{3}{4\pi}b g^{\prime}(b).
\end{equation}
To this effect we calculate
\begin{eqnarray}
g(a,b,\epsilon) & = & \mu(\Gamma(a,b,\epsilon)) \nonumber \\
    & = & \int\chi_{\Gamma(a,b,\epsilon)}(p+q)\,\Omega_m(dp)\,\Omega_m(dq) \nonumber \\
    & \approx & \int\chi_{(a,b)\times B_{\epsilon}({\vct 0})}(p+q)\,\Omega_m(dp)\,\Omega_m(dq) \nonumber \\
    & = & \int\chi_{(a,b)}(\omega_m({\vct p})+\omega_m({\vct q}))\chi_{B_{\epsilon}({\vct 0})}({\vct p}+{\vct q})\,\frac{d{\vct p}}{\omega_m({\vct p})}\frac{d{\vct q}}{\omega_m({\vct q})} \nonumber \\
    & = & \int\chi_{(a,b)}(\omega_m({\vct p})+\omega_m({\vct q}))\chi_{B_{\epsilon}({\vct 0})-{\vct q}}({\vct p})\,\frac{d{\vct p}}{\omega_m({\vct p})}\frac{d{\vct q}}{\omega_m({\vct q})} \nonumber \\
    & \approx & \int\chi_{(a,b)}(2\omega_m({\vct q}))\frac{\frac{4}{3}\pi\epsilon^3}{\omega_m({\vct q})^2}\,d{\vct q}. \nonumber
\end{eqnarray}
We will call this argument Argument 1. See Appendix 4 for a rigorous justification of Argument 1. Now 
\begin{eqnarray}
a<2\omega_m({\vct q})<b & \Leftrightarrow & \left(\frac{a}{2}\right)^2-m^2<{\vct q}^2<\left(\frac{b}{2}\right)^2-m^2 \nonumber \\
    & \Leftrightarrow & mZ(a)<|{\vct q}|<mZ(b), \nonumber
\end{eqnarray}
where
\begin{equation} \label{eq:Z_def}
Z(s)=\left(\frac{s^2}{4m^2}-1\right)^{\frac{1}{2}}\mbox{ for }s\geq2m.
\end{equation}
Thus
\begin{equation}
g(a,b,\epsilon)\approx\frac{16\pi^2}{3}\epsilon^3\int_{r=mZ(a)}^{mZ(b)}\frac{r^2}{m^2+r^2}\,dr.
\end{equation}
Hence
\begin{equation}
g_a(b)=\frac{16\pi^2}{3}\int_{r=mZ(a)}^{mZ(b)}\frac{r^2}{m^2+r^2}\,dr.
\end{equation}
Therefore $g_a$ is continuously differentiable and so Theorem~\ref{Theorem:Th2} applies.
Using the fundamental theorem of calculus
\begin{equation}
g_a^{\prime}(b)=\frac{16\pi^2}{3}\frac{m^2Z^2(b)}{m^2+m^2Z^2(b)}mZ^{\prime}(b)=\frac{16\pi^2}{3}\frac{mZ(b)}{b}.
\end{equation}
Therefore we compute the spectrum of $\mu$ as
\begin{equation}
\sigma(b)=4\pi mZ(b) \mbox{ for } b\ge2m.
\end{equation}

\section{Investigation of the measure defined by the convolution $\Omega_{im}*\Omega_{im}$ \label{section:imaginary_prop_conv_spec}}

The measure $\Omega^{+}_{im}$ is defined by
\begin{equation}
\Omega_{im}^{+}(\Gamma)=\int_{\pi(\Gamma\cap H_{im}^{+})}\frac{d{\vct p}}{\omega_{im}({\vct p})} \mbox{ for }\Gamma\in{\mathcal B}_0({\bf R}^4),
\end{equation}
where
\begin{equation}
H_{im}^{+}=\{p\in{\bf R}^4:p^2=-m^2, p^0\geq0\}.
\end{equation}
$\Omega_{im}^{+}$ is a measure concentrated on the positive time imaginary mass hyperboloid $H_{im}^{+}$ corresponding to mass $im$. There is also a measure $\Omega_{im}^{-}$ on $H_{im}^{-}$ and we may define $\Omega_{im}=\Omega_{im}^{+}+\Omega_{im}^{-}$, for~$m>0$. $\Omega_{im}$ is a Lorentz invariant measure on $H_{im}=\{p\in{\bf R}^4:p^2=-m^2\}$.

Define, for~$m\in{\bf C}$
\begin{equation}
J_m^{+}=\{p\in{\bf C}^4:p^2=m^2,\mbox{Re}(p^0)\geq0, \mbox{Im}(p^0)\geq0\},
\end{equation}
where $p^2=\eta_{\mu\nu}p^{\mu}p^{\nu}$ (in which $\eta_{\mu\nu}$ is the Minkowski space metric tensor). Then, for~$m>0$,
\begin{equation}
J_m^{+}\cap{\bf R}^4=\{p\in{\bf R}^4:p^2=m^2, p^0\geq0\}=H_m^{+},
\end{equation}
\begin{eqnarray} \label{eq:cx_hyp_property}
J_m^{+}\cap(i{\bf R}^4) & = & \{p\in i{\bf R}^4:p^2=m^2, \mbox{Re}(p^0)\geq0, \mbox{Im}(p^0)\geq0\} \nonumber \\
    & = & \{iq:q\in{\bf R}^4, q^2=-m^2, q^0\geq0\} \nonumber \\
    & = & iH_{im}^{+}. 
\end{eqnarray}

Now if ${\vct p}\in{\bf R}^3, m>0$, we may write
\[ \omega_{im}({\vct p})=((im)^2+{\vct p}^2)^{\frac{1}{2}}=(-m^2+{\vct p}^2)^{\frac{1}{2}}=(-(m^2+(i{\vct p})^2))^{\frac{1}{2}}=i(m^2+(i{\vct p})^2)^{\frac{1}{2}}=i\omega_m(i{\vct p}). \] 

One may consider the measure $\Omega_m^{+}$ to be defined on $i{\bf R}^4$ as well as ${\bf R}^4$ and for all $m\in{\bf R}$ or $m\in i{\bf R}$ according to
\begin{equation}
\Omega_{m}^{+}({\Gamma})=\int_{\pi(\Gamma\cap J_{m}^{+})}\frac{d{\vct p}}{\omega_{m}({\vct p})}.
\end{equation}
Then from Equation~(\ref{eq:cx_hyp_property})
\begin{equation}
\Omega_m^{+}(i\Gamma)=\int_{i\pi(\Gamma\cap H_{im}^{+})}\frac{d{\vct p}}{\omega_m({\vct p})}.
\end{equation}
Now make the substitution ${\vct p}=i{\vct q}$. Then $d{\vct p}=-id{\vct q}$.  
Thus
\begin{equation}
\Omega_m^{+}(i\Gamma)=\int_{\pi(\Gamma\cap H_{im}^{+})}\frac{-id{\vct q}}{-i\omega_{im}({\vct q})}=\Omega_{im}^{+}(\Gamma). 
\end{equation}
Now suppose that
\begin{equation}
\psi=\sum_k c_k\chi_{E_k},
\end{equation}
where $c_i\in{\bf C}$ and $E_k\in{\mathcal B}_0({\bf R}^4)$, is a simple function. Then 
\begin{equation}
\begin{aligned}
\int_{{\bf R}^4}\psi(p)\,\Omega_{im}^{+}(dp)  = &~\sum_k c_k\Omega_{im}^{+}(E_k)  \\
     = &~\sum_k c_k\Omega_m^{+}(iE_k)  \\
     = &~\sum_k c_k\int_{i{\bf R}^4}\chi_{iE_k}(p)\,\Omega_m^{+}(dp)  \\
     = &~\sum_k c_k\int_{i{\bf R}^4}\chi_{E_k}(\frac{p}{i})\,\Omega_m^{+}(dp)  \\
     = &~\int_{i{\bf R}^4}\psi(\frac{p}{i})\,\Omega_m^{+}(dp). \\
\end{aligned}
\end{equation}
As this is true for every such simple function $\psi$ it follows that
\begin{equation}
\int_{{\bf R}^4}\psi(p)\,\Omega_{im}^{+}(dp)=\int_{i{\bf R}^4}\psi(\frac{p}{i})\,\Omega_m^{+}(dp),
\end{equation}
for every function $\psi$ which is integrable with respect to $\Omega_{im}^{+}$. Therefore
\begin{equation}
\begin{aligned} \label{eq:convolve}
(\Omega_{im}^{+}*\Omega_{im}^{+})(\Gamma)  = &~\int_{({\bf R}^4)^2}\chi_{\Gamma}(p+q)\,\Omega_{im}^{+}(dp)\,\Omega_{im}^{+}(dq)  \\
     = &~\int_{(i{\bf R}^4)^2}\chi_{\Gamma}\left(\frac{p+q}{i}\right)\,\Omega_m^{+}(dp)\,\Omega_m^{+}(dq)  \\
     = &~\int_{(i{\bf R}^4)^2}\chi_{i\Gamma}(p+q)\Omega_m^{+}(dp)\Omega_m^{+}(dq)  \\
     = &~(\Omega_m^{+}*\Omega_m^{+})(i\Gamma),
\end{aligned}
\end{equation}
for all $\Gamma\in{\mathcal B}_0({\bf R}^4)$.

Now in general, suppose that a measure $\mu$ has a causal spectral representation of the form
\begin{equation}
\mu(\Gamma)=\int_{m^{\prime}=0}^{\infty}\Omega_{m^{\prime}}^{+}(\Gamma)\,\sigma(m^{\prime}),
\end{equation}
for some spectral measure $\sigma:{\mathcal B}_0([0,\infty))\rightarrow{\bf C}$. 
Then $\mu$ extends to a measure  defined on $i{\bf R}^4$ by
\begin{equation}
\mu(i\Gamma)=\int_{m^{\prime}=0}^{\infty}\Omega_{m^{\prime}}^{+}(i\Gamma)\,\sigma(dm^{\prime})=\int_{m^{\prime}=0}^{\infty}\Omega_{im^{\prime}}^{+}(\Gamma)\,\sigma(dm^{\prime}),
\end{equation}
for $\Gamma\in{\mathcal B}_0({\bf R}^4)$.
Therefore since, as~we have determined above, $\Omega_m^{+}*\Omega_m^{+}$ is a causal spectral measure with spectrum
\begin{equation} \label{eq:imaginary_spectrum}
\sigma(m^{\prime})=\left\{\begin{array}{l}
4\pi mZ(m^{\prime}) \mbox{ for } m^{\prime}\ge2m \\
0\mbox{ otherwise},\end{array}\right.
\end{equation}
it follows that
\begin{equation}
(\Omega_m^{+}*\Omega_m^{+})(i\Gamma)=\int_{m^{\prime}=0}^{\infty}\Omega_{im^{\prime}}^{+}(\Gamma)\,\sigma(dm^{\prime}).
\end{equation}
Therefore using Equation~(\ref{eq:convolve}) $\Omega_{im}^{+}*\Omega_{im}^{+}$ is a measure with spectral representation
\begin{equation}
(\Omega_{im}^{+}*\Omega_{im}^{+})(\Gamma)=\int_{m^{\prime}=0}^{\infty}\Omega_{im^{\prime}}^{+}(\Gamma)\,\sigma(m^{\prime})\,dm^{\prime},
\end{equation}
where $\sigma$ is the spectral function given by Equation~(\ref{eq:imaginary_spectrum}).
 Note that $\Omega_{im}^{+}*\Omega_{im}^{+}$ is not causal, it is a type III measure, and~\begin{equation}
\mbox{supp}(\Omega_{im}^{+}*\Omega_{im}^{+})=\{p\in{\bf R}^4:p^2\leq-4m^2, p^0\geq0\}.
\end{equation}

\section{Determination of the density defining a causal Lorentz invariant measure from its spectrum\label{section:density}}

Suppose that $\mu$ is of the form of Eq.~\ref{eq:invariant2} where $\sigma$ is a well behaved (e.g. locally integrable) function. We would like to see if $\mu$ can be defined by a density with respect to the Lebesgue measure, i.e. if there exists a function $f:{\bf R}^4\rightarrow{\bf C}$ such that
\begin{equation}
\mu(\Gamma)=\int_{\Gamma}f(p)\,dp.
\end{equation}
Well we have that
\begin{equation}
\mu(\Gamma)=\int_{m=0}^{\infty}\sigma(m)\Omega_m^{+}(\Gamma)\,dm=\int_{m=0}^{\infty}\sigma(m)\int_{\pi(\Gamma\cap  H_m^{+})}\frac{d{\vct p}}{\omega_m({\vct p})}\,dm.
\end{equation}
Now 
\begin{eqnarray}
{\vct p}\in\pi(\Gamma\cap H_m^{+}) & \Leftrightarrow & (\exists p\in{\bf R}^4){\vct p}=\pi(p),p\in H_m^{+}, p\in\Gamma \nonumber \\
    & \Leftrightarrow & (\omega_m({\vct p}),{\vct p})\in\Gamma \nonumber \\
    & \Leftrightarrow & \chi_{\Gamma}(\omega_m({\vct p}),{\vct p})=1. \nonumber
\end{eqnarray}
Therefore
\begin{equation}
\mu(\Gamma)=\int_{m=0}^{\infty}\sigma(m)\int_{{\bf R}^3}\chi_{\Gamma}(\omega_m({\vct p}),{\vct p})\frac{1}{\omega_m({\vct p})}\,d{\vct p}\,dm.
\end{equation}
Now consider the transformation defined by the function $h:(0,\infty)\times{\bf R}^3\rightarrow{\bf R}^4$ given by
\begin{equation}
h(m,{\vct p})=(\omega_m({\vct p}),{\vct p}).
\end{equation}
Let 
\begin{equation}
q=h(m,{\vct p})=(\omega_m({\vct p}),{\vct p})=((m^2+{\vct p}^2)^{\frac{1}{2}},{\vct p}).
\end{equation}
Then
\begin{equation}
\frac{\partial q^0}{\partial m}=m\omega_m({\vct p})^{-1}, \frac{\partial q^0}{\partial p^j}= p^j\omega_m({\vct p})^{-1},  \frac{\partial q^i}{\partial m}=0, \frac{\partial q^i}{\partial p^j}=\delta_{ij}, 
\end{equation}
for $i,j=1,2,3$. Thus the Jacobian of the transformation is
\begin{equation}
J(m,{\vct p})=m\omega_m({\vct p})^{-1}.
\end{equation}
Now $q=(\omega_m({\vct p}),{\vct p})$. Therefore $q^2=\omega_m({\vct p})^2-{\vct p}^2=m^2$. So $m=(q^2)^{\frac{1}{2}},q^2>0$.
Thus
\begin{eqnarray}
\mu(\Gamma) & = & \int_{q\in{\bf R}^4,q^2>0,q^0>0}\chi_{\Gamma}(q)\frac{\sigma(m)}{\omega_m({\vct p})}\frac{dq}{J(m,{\vct p})} \nonumber \\
    & = & \int_{q^2>0,q^0>0}\chi_{\Gamma}(q)\frac{\sigma(m)}{m}\,dq.
\end{eqnarray}
Hence
\begin{eqnarray}
\mu(\Gamma) & = & \int_{q^2>0,q^0>0}\chi_{\Gamma}(q)\frac{\sigma((q^2)^{\frac{1}{2}})}{(q^2)^{\frac{1}{2}}}\,dq \nonumber \\
    & = & \int_{\Gamma}f(q)\,dq, \nonumber
\end{eqnarray}
where $f:{\bf R}^4\rightarrow{\bf C}$ is defined by
\begin{equation} \label{eq:equiv_density}
f(q)=\left\{\begin{array}{l}
(q^2)^{-\frac{1}{2}}\sigma((q^2)^{\frac{1}{2}})\mbox{ if } q^2>0,q^0>0 \\
0\mbox{ otherwise.}
\end{array}\right.
\end{equation}
We have therefore shown how, given a spectral representation of a causal measure in which the spectrum is a complex function, one can obtain and equivalent representation of the measure in terms of a density with respect to Lebesgue measure.

\section{Lorentz covariant tensor valued measures\label{section:tensor}}

A tensor valued measure $\mu^{\alpha\beta}:{\mathcal B}_0({\bf R}^4)\rightarrow{\bf C}$ will be said to be Lorentz covariant if
\begin{equation}
\mu^{\alpha\beta}(\Lambda\Gamma)={\Lambda^{\alpha}}_{\rho}{\Lambda^{\beta}}_{\sigma}\mu^{\rho\sigma}(\Gamma),\forall\Lambda\in O(1,3),\Gamma\in{\mathcal B}_0({\bf R}^4).
\end{equation}
There is a general class of causal Lorentz covariant 2-tensor valued measures of the form
\begin{equation}
\mu^{\alpha\beta}(\Gamma)=\int_{m=0}^{\infty}\int_{\Gamma}\eta^{\alpha\beta}p^2\,\Omega_m(dp)\,\sigma_1(dm)+\int_{m=0}^{\infty}\int_{\Gamma}p^{\alpha}p^{\beta}\,\Omega_m(dp)\,\sigma_2(dm), \label{eq:general_L_i_measure}
\end{equation}
for spectral measures $\sigma_1,\sigma_2:{\mathcal B}_0((0,\infty))\rightarrow{\bf C}$. If $\mu^{\alpha\beta}$ is a causal Lorentz covariant 2-tensor valued measure. Define
\[ g^{\alpha\beta}(a,b,\epsilon)=\mu^{\alpha\beta}(\Gamma(a,b,\epsilon))\mbox{ for }0<a<b,\epsilon>0. \] 
Then it can be shown using a generalization of the spectral calculus that $\mu^{\alpha\beta}$ has the form of Eq.~\ref{eq:general_L_i_measure} with spectral measures defined by continuous functions  if and only if 
\begin{align}
g^{\alpha\beta}_a(b)&=\lim_{\epsilon\rightarrow0}\epsilon^{-3}g^{\alpha\beta}(a,b,\epsilon),
\end{align}
exists and is continuously differentiable, in which case the spectral functions $\sigma_1$ and $\sigma_2$ can be determined using
\begin{align}
\sigma_1(b)&=-\frac{3}{4\pi}\frac{1}{b}g_a^{ii\prime}(b) \label{eq:tensor_1},i\in\{1,2,3\},\\
\sigma_2(b)&=\frac{3}{4\pi}\frac{1}{b}g_a^{00\prime}(b)-\sigma_1(b). \label{eq:tensor_2}
\end{align}

\section{Convolutions and products of causal Lorentz invariant measures \label{section:conv_and_products}}
\unskip
\subsection{Convolution of measures}

Let $\mu$ and $\nu$ be causal Lorentz invariant complex measures. Then (up to possible atoms at the origin which can be dealt with in a straightforward way) there exist spectral measures $\sigma,\rho:{\mathcal B}_0([0,\infty))\rightarrow{\bf C}$ such that
\begin{equation}
\begin{aligned}
\mu  = &~\int_{m=0}^{\infty}\Omega_m\,\sigma(dm),  \\
\nu  = &~\int_{m=0}^{\infty}\Omega_m\,\rho(dm).
\end{aligned}
\end{equation}
We will assume, without loss of generality, that $\sigma$ and $\rho$ are complex measures in the usual sense of the term, so $\sigma,\rho:{\mathcal B}([0,\infty))\rightarrow{\bf C}$. The convolution of $\mu$ and $\nu$, if~it exists, is given by
\begin{equation}
(\mu*\nu)(\Gamma)=\int\chi_{\Gamma}(p+q)\,\mu(dp)\,\nu(dq).
\end{equation}
Now let $\psi=\sum_i c_i\chi_{E_i}$ with $c_i\in{\bf C}, E_i\in{\mathcal B}_0({\bf R}^4)$ be a simple function. Then
\begin{eqnarray}
\int\psi(p)\,\mu(dp) & = & \int\sum_i c_i\chi_{E_i}\,\mu(dp) \nonumber \\
    & = & \sum_i c_i\mu(E_i) \nonumber \\
    & = & \sum_i c_i\int_{m=0}^{\infty}\Omega_m(E_i)\,\sigma(dm) \nonumber \\
    & = & \sum_i c_i\int_{m=0}^{\infty}\int_{{\bf R}^4}\chi_{E_i}(p)\,\Omega_m(dp)\,\sigma(dm) \nonumber \\
    & = & \int_{m=0}^{\infty}\int_{{\bf R}^4}\psi(p)\,\Omega_m(dp)\,\sigma(dm). \nonumber
\end{eqnarray}
Therefore for any sufficiently well behaved measurable function $\psi:{\bf R}^4\rightarrow{\bf C}$ (e.g. bounded measurable functions of compact support)
\begin{equation}
\int\psi(p)\mu(dp)=\int\psi(p)\,\Omega_m(dp)\,\sigma(dm).
\end{equation}
(Note that the integral exists because $\sigma$ is a (Borel) measure.) Hence for all $\Gamma\in{\mathcal B}_0({\bf R}^4)$
\begin{equation}
\begin{aligned} \label{eq:convolve1}
(\mu*\nu)(\Gamma)  = &~\int\chi_{\Gamma}(p+q)\,\mu(dp)\,\nu(dq)  \\
     = &~\int\chi_{\Gamma}(p+q)\,\Omega_m(dp)\,\sigma(dm)\,\Omega_{m^{\prime}}(dq)\,\rho(dm^{\prime})  \\
     = &~\int\chi_{\Gamma}(p+q)\,\Omega_m(dp)\,\Omega_{m^{\prime}}(dq)\sigma(dm)\,\rho(dm^{\prime}),  
\end{aligned}
\end{equation}
by Fubini's theorem, as~long as
\begin{equation}
\int\chi_{\Gamma}(p+q)\,\Omega_m(dp)\,\Omega_{m^{\prime}}(dq)|\sigma|(dm)<\infty,\forall m^{\prime}\in[0,\infty),
\end{equation}
where $|\sigma|$ is the total variations of the measure $\sigma$. 

Suppose that $\Gamma\in{\mathcal B}_0({\bf R}^4)$. Then there exists $a,R\in(0,\infty)$ such that $\Gamma\subset(-a,a)\times B_R({\vct 0})$, where~$B_R({\vct 0})=\{{\vct p}\in{\bf R}^3:|{\vct p}|<R\}$. Now
\begin{equation}
\int\chi_{\Gamma}(p+q)\,\Omega_m(dp)=\int_{\Gamma-q}\Omega_m(dp)=\Omega_m(\Gamma-q)<\infty,
\end{equation}
for all $q\in{\bf R}^4$ because $\Omega_m$ is Borel and $\Gamma$ is relatively compact. 

Now suppose that $m,m^{\prime}>a$. Then
\begin{equation}
p\in H_m^{+},q\in H_{m^{\prime}}^{+}\Rightarrow(p+q)^0=p^0+q^0\geq m+m^{\prime}>2a\Rightarrow (p+q){\notin}\Gamma.
\end{equation}
Thus
\begin{equation}
\int\chi_{\Gamma}(p+q)\,\Omega_m(dp)\,\Omega_{m^{\prime}}(dq)=0.
\end{equation}

Therefore since $\sigma$ and $\rho$ are Borel, $(\mu*\nu)(\Gamma)$ exists, is finite and is given by Equation~(\ref{eq:convolve1}). 

Now let $\Lambda\in O(1,3)^{+\uparrow}$, $\psi:{\bf R}^4\rightarrow{\bf C}$ be a measurable function of compact support. Then
\begin{eqnarray}
<\mu*\nu,\Lambda\psi> & = & \int\psi(\Lambda^{-1}(p+q))\,\Omega_m(dp)\,\Omega_{m^{\prime}}(dq)\,\sigma(dm)\,\rho(dm^{\prime}) \nonumber \\
    & = & \int\psi(p+q)\,\Omega_m(dp)\,\Omega_{m^{\prime}}(dq)\,\sigma(dm)\,\rho(dm^{\prime}). \nonumber \\
    & = & <\mu*\nu,\psi> \nonumber
\end{eqnarray}
Therefore $\mu*\nu$ is Lorentz invariant. It can be shown, by~an argument similar to that used for the case $\Omega_m*\Omega_m$ that $\mu*\nu$ is~causal.

We have therefore shown that the convolution of two causal Lorentz invariant complex measures exists and is a causal Lorentz invariant complex~measure.

\subsection{Product of~measures}

We now turn to the problem of computing the product of two causal Lorentz invariant complex measures. The~problem of computing the product of measures or distributions is difficult in general and has attracted a large amount of research. In such work one generally seeks a definition of the product of measures or distributions which agrees with the ordinary product when the measures or distributions are functions (i.e., densities with respect to Lebesgue measure). The~most common approach is to use the fact that, for~Schwartz functions $f,g\in{\mathcal S}({\bf R}^4)$ multiplication in the spatial domain corresponds to convolution in the frequency domain, i.e.,~$(fg)^{\wedge}=f^{\wedge}*g^{\wedge}$ (where $\wedge$ denotes the Fourier transform operator). Thus one defines the product of measures or distributions $\mu,\nu$ as
\begin{equation}
\mu\nu=(\mu^{\wedge}*\nu^{\wedge})^{\vee}.
\end{equation}
However, this definition is only successful when the convolution that it involves exists which may not be the case in general. If~$\mu,\nu$ are tempered measures then $\mu^{\wedge}$ and $\nu^{\wedge}$ exist as tempered distributions, however, they are generally not causal, even if $\mu,\nu$ are~causal. 

We will therefore not use the ``frequency space" approach to define the product of measures but will use a different approach. Our approach is just as valid as the frequency space approach because our product will coincide with the usual function product when the measures are defined by densities. Furthermore, our approach is useful for the requirements of QFT because measures and distributions in QFT are frequently Lorentz invariant and~causal.

Let int$(C)=\{p\in{\bf R}^4:p^2>0,p^0>0\}$. Suppose that $f:\mbox{int}(C)\rightarrow{\bf C}$ is a Lorentz invariant locally integrable function. Then it defines a causal Lorentz invariant measure $\mu_f$ which, by~the spectral theorem, must have a representation of the form
\begin{equation}
\mu_f(\Gamma)=\int_{\Gamma}f(p)\,dp=\int_{m=0}^{\infty}\Omega_m(\Gamma)\,\sigma(dm),
\end{equation}
for some spectral measure $\sigma:{\mathcal B}_0([0,\infty))\rightarrow{\bf C}$. As $\mu_f$ is absolutely continuous with respect to Lebesgue measure it follows that $\sigma$ must be non-singular, i.e.,~a function.
By the result of the previous section a density defining $\mu_f$ is ${\tld f}:\mbox{int}(C)\rightarrow{\bf C}$ defined by
\begin{equation}
{\tld f}(p)=(p^2)^{-\frac{1}{2}}\sigma((p^2)^{\frac{1}{2}}), p\in\mbox{ int}(C).
\end{equation}
We must have that ${\tld f}=f$ (almost everywhere). Therefore (almost everywhere on int$(C)$)
\begin{equation}
f(p)=(p^2)^{-\frac{1}{2}}\sigma((p^2)^{\frac{1}{2}}). \label{equation:f_from_sigma}
\end{equation}
Without loss of generality, it can be assumed that equality holds everywhere in Equation~(\ref{equation:f_from_sigma}). 
$f(p)$~depends only on $p^2$. Therefore for all $m>0$, $\sigma(m)=mf(p)$ for all $p\in\mbox{int}(C)$ such that $p^2=m^2$. In~particular
\begin{equation}
\sigma(m)=mf((m,{\vct 0})^{T}), \forall m>0.
\end{equation}
Now we are seeking a definition of product which has useful properties. Two such properties would be that it is distributive with respect to generalized sums such as integrals and also that it agrees with the ordinary product when the measures are defined by functions. Suppose that we had such a product. Let $f,g:\mbox{int}(C)\rightarrow{\bf C}$ be Lorentz invariant locallly integrable functions. Let $\mu,\nu:{\mathcal B}_0(\mbox{int}(C))\rightarrow{\bf C}$ be the associated measures with spectra $\sigma,\rho$. Then
\begin{eqnarray}
\mu\nu & = & \int_{m=0}^{\infty}\Omega_m\,\sigma(dm)\int_{m^{\prime}=0}^{\infty}\Omega_{m^{\prime}}\,\rho(dm^{\prime}) \nonumber \\
    & = & \int_{m=0}^{\infty}\Omega_m\,mf((m,{\vct 0})^{T})\,dm\int_{m^{\prime}=0}^{\infty}\Omega_{m^{\prime}}\,m^{\prime}g((m^{\prime},{\vct 0})^{T})\,dm^{\prime} \nonumber \\
    & = & \int_{m=0}^{\infty}\int_{m^{\prime}=0}^{\infty}\Omega_m\Omega_{m^{\prime}}mf((m,{\vct 0})^{T})m^{\prime}g((m^{\prime},{\vct 0})^{T})\,dm\,dm^{\prime}. \nonumber 
\end{eqnarray}
Now we want this to be equal to
\begin{equation}
\int_{m=0}^{\infty}\Omega_mm(fg)((m,{\vct 0})^{T})\,dm
\end{equation}
This will be the case (formally) if we have
\begin{equation}
\Omega_m\Omega_{m^{\prime}}=\frac{1}{m}\delta(m-m^{\prime})\Omega_m, \forall m,m^{\prime}>0.
\end{equation}
Physicists will be familiar with such a formula (e.g., the equal time commutation relations). Rather than attempting to define its meaning in a rigorous way, we will simply carry out the following formal computation for general Lorentz invariant measures $\mu,\nu$ with spectra $\sigma,\rho$
\begin{eqnarray}
\mu\nu & = & \int_{m=0}^{\infty}\Omega_m\,\sigma(dm)\int_{m^{\prime}=0}^{\infty}\Omega_{m^{\prime}}\,\rho(dm^{\prime}) \nonumber \\
    & = & \int_{m=0}^{\infty}\int_{m^{\prime}=0}^{\infty}\Omega_m\Omega_{m^{\prime}}\sigma(m)\rho(m^{\prime})\,dm\,dm^{\prime} \nonumber \\
    & = & \int_{m=0}^{\infty}\int_{m^{\prime}=0}^{\infty}\frac{1}{m}\Omega_m\delta(m-m^{\prime})\sigma(m)\rho(m^{\prime})\,dm^{\prime}\,dm \nonumber \\
    & = & \int_{m=0}^{\infty}\frac{1}{m}\Omega_m\sigma(m)\rho(m)\,dm. \nonumber
\end{eqnarray}
Therefore we can simply define the product $\mu\nu$ in general by
\begin{equation}
\mu\nu=\int_{m=0}^{\infty}\frac{1}{m}\Omega_m\,(\sigma\rho)(dm),
\end{equation}
i.e.,
\begin{equation}
(\mu\nu)(\Gamma)=\int_{m=0}^{\infty}\frac{1}{m}\Omega_m(\Gamma)\,(\sigma\rho)(dm),
\end{equation}
for $\Gamma\in{\mathcal B}_0({\bf R}^4)$.

We have therefore reduced the problem of computing the product of measures on int$(C)$ to the problem of computing the product of their 1D spectral measures. The~problem of multiplying 1D measures is somewhat less problematic than the problem of multiplying 4D measures. A~large class of 1D measures is made up of measures which are of the form of a function plus a finite number of ``atoms'' (singularities of the form $c\delta_a$ where $c\in{\bf C}\backslash\{0\},a\in[0,\infty)$, where $\delta_a$ is the Dirac delta function (measure) concentrated at $a$). There are other pathological types of the 1D measure but these may not be of interest for physical~applications. 

In the general non-pathological case, if~$\mu,\nu$ are causal Lorentz invariant measures with spectra $\sigma(m)=\xi(m)+\sum_{i=1}^{k}c_i\delta(m-a_i), \rho(m)=\zeta(m)+\sum_{j=1}^l d_j\delta(m-b_j)$ where $\xi,\zeta:[0,\infty)\rightarrow{\bf C}$ are locally integrable functions, $k,l\geq0,c_i,d_j\in{\bf C}\backslash\{0\},a_i,b_j\in[0,\infty)$ are such that $a_i\neq b_j,\forall i,j$ then we may define the product of $\mu$ and $\nu$ to be the causal Lorentz invariant measure $\mu\nu$ given by
\begin{equation}
\mu\nu=\int_{m=0}^{\infty}\Omega_m\tau(dm),
\end{equation}
where
\begin{eqnarray}
\tau(m) & = & \frac{1}{m}(\xi(m)\zeta(m)+\zeta(m)\sum_{i=1}^kc_i\delta(m-a_i)+\xi(m)\sum_{j=1}^ld_j\delta(m-b_j)) \nonumber \\
 & = & \frac{1}{m}(\xi(m)\zeta(m)+\sum_{i=1}^k\zeta(a_j)c_i\delta(m-a_i)+\sum_{j=1}^l\xi(b_j)d_j\delta(m-b_j)), \nonumber
\end{eqnarray}
for $m>0$.
\section{Vacuum polarization\label{section:vacuum_polarization}}

\subsection{Definition of the vacuum polarization tensor as a Lorentz covariant tempered complex tensor valued measure $\Pi^{\mu\nu}$}

The vacuum polarization tensor is given by
\begin{align}
\Pi^{\mu\nu}(k)&=-\mbox{Tr}(\int\frac{dp}{(2\pi)^{4}}i(-e)\gamma^{\mu}iS(p)i(-e)\gamma^{\nu}iS(p-k))\label{eq:vac_pol_def}\\
&=-\frac{e^2}{(2\pi)^{4}}\int\mbox{Tr}(\gamma^{\mu}\frac{1}{{\slas p}-m+i\epsilon}\gamma^{\nu}\frac{1}{{\slas p}-{\slas k}-m+i\epsilon})\,dp,\label{eq:vp_tensor}
\end{align}
where $e$ is the magnitude of the charge of the electron, $m$ is the mass of the electron and $S$ is the electron propagator. (The first minus sign in Eq.~\ref{eq:vac_pol_def} is associated with the fermion loop). 
This can be rewritten as
\begin{equation} \label{eq:vp_tensor_1}
\Pi^{\mu\nu}(k)=-\frac{e^2}{(2\pi)^4}\int\frac{\mbox{Tr}(\gamma^{\mu}({\slas p}+m)\gamma^{\nu}({\slas p}-{\slas k}+m))}{(p^2-m^2+i\epsilon)((p-k)^2-m^2+i\epsilon)}\,dp.
\end{equation}
As is well known, the integral defining this ``function" is divergent for all $k\in{\bf R}^4$ and all the machinery of regularization and renormalization has been developed to get around this problem.

We propose that the object defined by Eq.~\ref{eq:vp_tensor_1} exists when viewed as a tensor valued complex measure on Minkowski space. To show this, suppose that $\Pi^{\mu\nu}$ were a tensor valued complex measure which is defined by a density which we also denote as $\Pi^{\mu\nu}$. Then we may make the following formal computation.
\begin{eqnarray}
\Pi^{\mu\nu}(\Gamma) & = & \int_{\Gamma}\Pi^{\mu\nu}(k)\,dk \nonumber \\
    & = & \int\chi_{\Gamma}(k)\Pi^{\mu\nu}(k)\,dk \nonumber \\
    & = & -\frac{e^2}{(2\pi)^4}\int\chi_{\Gamma}(k)\frac{\mbox{Tr}(\gamma^{\mu}({\slas p}+m)\gamma^{\nu}({\slas p}-{\slas k}+m))}{(p^2-m^2+i\epsilon)((p-k)^2-m^2+i\epsilon)}\,dp\,dk \nonumber \\
   & = & -\frac{e^2}{(2\pi)^4}\int\chi_{\Gamma}(k)\frac{\mbox{Tr}(\gamma^{\mu}({\slas p}+m)\gamma^{\nu}({\slas p}-{\slas k}+m))}{(p^2-m^2+i\epsilon)((p-k)^2-m^2+i\epsilon)}\,dk\,dp \nonumber \\ 
   & = & -\frac{e^2}{(2\pi)^4}\int\chi_{\Gamma}(k+p)\frac{\mbox{Tr}(\gamma^{\mu}({\slas p}+m)\gamma^{\nu}({-\slas k}+m))}{(p^2-m^2+i\epsilon)(k^2-m^2+i\epsilon)}\,dk\,dp \nonumber \\
   & = & \frac{e^2}{(2\pi)^4}\int\chi_{\Gamma}(k+p)\frac{\mbox{Tr}(\gamma^{\mu}({\slas p}+m)\gamma^{\nu}({\slas k}-m))}{(p^2-m^2+i\epsilon)(k^2-m^2+i\epsilon)}\,dk\,dp \nonumber 
\end{eqnarray}
(where $\Gamma$ is a well behaved, e.g. relatively compact Borel, subset of ${\bf R}^4$).

Now the propagators in QFT can be viewed in a rigorous fashion as measures on Minkowski space and we make the identification or {\em ansatz}
\begin{equation}
\frac{1}{p^2-m^2+i\epsilon}\rightarrow -\pi i\Omega_m^{\pm}(p), m\ge0,
\end{equation}
where $\Omega_m^{\pm}$ is the standard Lorentz invariant measure on the mass shell hyperboloid (cone) $H_m^{\pm}$ corresponding to mass $m>0$ ($m=0$) (see (Mashford, 2017b) for explanation). Therefore the outcome of our formal computation is that
\begin{equation} \label{eq:Pi_gen}
\Pi^{\mu\nu}(\Gamma)=-\frac{e^2}{16\pi^2}\int\chi_{\Gamma}(k+p)\mbox{Tr}(\gamma^{\mu}({\slas p}+m)\gamma^{\nu}({\slas k}-m))\,\Omega_m^{\pm}(dk)\,\Omega_m^{\pm}(dp).
\end{equation}
We will consider the $++$ case corresponding to a virtual electron positron pair. (The $--$ case corresponds to a virtual positron electron pair. The $+-$ and $-+$ cases are divergent). Thus
\begin{equation} \label{eq:Pi2}
\Pi^{\mu\nu}(\Gamma)=-\frac{e^2}{16\pi^2}\int\chi_{\Gamma}(k+p)\mbox{Tr}(\gamma^{\mu}({\slas p}+m)\gamma^{\nu}({\slas k}-m))\,\Omega_m(dk)\,\Omega_m(dp), m>0.
\end{equation}
(We use the symbol $\Omega_m$ to denote $\Omega_m^{+}$ if $m>0$ or $\Omega_{|m|}^{-}$ if $m<0$.)

The important thing is that the object defined by Eq.~\ref{eq:Pi2} exists as a tensor valued complex measure (i.e. when its argument is a relatively compact Borel subset of ${\bf R}^4$). This is because
\begin{equation}
\int\chi_{\Gamma}(k+p)|\mbox{Tr}(\gamma^{\mu}({\slas p}+m)\gamma^{\nu}({\slas k}-m))|\,\Omega_m(dk)\,\Omega_m(dp)<\infty, 
\end{equation}
for all $\Gamma\in{\mathcal B}_0({\bf R}^4)$. It also exists as a tempered complex tensor valued distribution since
\begin{equation} \label{eq:Pi3}
\int\psi(k+p)\mbox{Tr}(\gamma^{\mu}({\slas p}+m)\gamma^{\nu}({\slas k}-m))\,\Omega_m(dk)\,\Omega_m(dp),
\end{equation}
is convergent for any Schwartz function $\psi\in{\mathcal S}({\bf R}^4,{\bf C})$. The basic reason for both these facts is that as $|k|,|p|\rightarrow{\infty}$ with $k,p\in\mbox{ the mass shell } H_m^{+}$, $(k+p)^0\rightarrow{\infty}$. Hence $(k+p)$ is eventually ${\slas\in}\Gamma$ for any relatively compact $\Gamma$ and $\psi(k+p)\rightarrow0$ rapidly for any Schwartz function $\psi\in{\mathcal S}({\bf R}^4,{\bf C})$.

Thus $\Pi^{\mu\nu}$ exists as a tempered measure. Therefore we have in a few lines of formal argument arrived at an object which has a well defined existence and can investigate the properties of this object $\Pi$ without any further concern about ill-definedness or the fear of propagating ill-definedness through our calculations.

Define $\Pi:{\mathcal B}_0({\bf R}^4)\rightarrow{\bf C}$ by
\begin{equation}
\Pi(\Gamma)=\eta_{\mu\nu}\Pi^{\mu\nu}(\Gamma),
\end{equation}
where $\eta=\mbox{diag}(1,-1,-1,-1)$ is the Minkowski space metric tensor. It is straightforward to show that the measure $\Pi$ is Lorentz invariant.

Using an argument similar to that for $\Omega_m*\Omega_m$ it can be shown that the both $\Pi^{\mu\nu}$ and $\Pi$ are supported in  $C_{2m}=\{p\in{\bf R}^4:p^2\geq4m^2,p^0>0\}$. 

\subsection{Application of the spectral calculus to determine the spectrum of the contracted vacuum polarization tensor $\Pi$}

We have shown that $\Pi$ is a Lorentz invariant tempered complex measure with support contained in $C_{2m}$. Therefore by the spectral theorem  $\Pi$ must have a spectral representation of the form
\begin{equation} \label{eq:vp_spectral1}
\Pi(\Gamma)=\int_{m^{\prime}=2m}^{\infty}\sigma(dm^{\prime})\Omega_{m^{\prime}}(\Gamma).
\end{equation}
We would like to compute the spectral measure $\sigma$. First we have
\begin{eqnarray}
&  & \mbox{Tr}(\eta_{\mu\nu}\gamma^{\mu}({\slas p}+m)\gamma^{\nu}({\slas k}-m)) \nonumber \\
&  & = \eta_{\mu\nu}p_{\alpha}k_{\beta}\mbox{Tr}(\gamma^{\mu}\gamma^{\alpha}\gamma^{\nu}\gamma^{\beta})+0+0-\mbox{Tr}(\gamma^{\mu}\gamma_{\mu}m^2) \nonumber \\
& & = \eta_{\mu\nu}p_{\alpha}k_{\beta}\mbox{Tr}(\gamma^{\mu}\gamma^{\alpha}\gamma^{\nu}\gamma^{\beta})-16m^2 \nonumber \\
& & = 4p_{\alpha}k_{\beta}\eta_{\mu\nu}(\eta^{\mu\alpha}\eta^{\nu\beta}-\eta^{\mu\nu}\eta^{\alpha\beta}+\eta^{\mu\beta}\eta^{\alpha\nu})-16m^2 \nonumber \\
& & = 4\eta_{\mu\nu}(p^{\mu}k^{\nu}-\eta^{\mu\nu}p.k+k^{\mu}p^{\nu})-16m^2 \nonumber \\
& & = 4(p.k-4p.k+p.k-4m^2) \nonumber \\
& & = -8(p.k+2m^2), \nonumber 
\end{eqnarray}
where we have used in the second line the fact that the trace of a product of an odd number of gamma matrices vanishes.

We now compute in a fashion similar to that used when determining the spectrum of $\Omega_m*\Omega_m$ in as follows (this computation is very intuitively plausible and can be justified rigorously in a fashion similar to the justification of Argument 1 (Appendix 4).
\begin{eqnarray}
g(a,b,\epsilon) & = & \Pi(\Gamma(a,b,\epsilon)) \nonumber \\
    & = & -\frac{e^2}{16\pi^2}\int\chi_{\Gamma(a,b,\epsilon)}(k+p)\mbox{Tr}(\eta_{\mu\nu}\gamma^{\mu}({\slas p}+m)\gamma^{\nu}({\slas k}-m))\,\Omega_m(dk)\,\Omega_m(dp) \nonumber \\
    & = &  \frac{e^2}{2\pi^2}\int\chi_{\Gamma(a,b,\epsilon)}(k+p)(p.k+2m^2)\,\Omega_m(dk)\,\Omega_m(dp) \nonumber \\
    & \approx & \frac{e^2}{2\pi^2}\int\chi_{(a,b)}(\omega_m({\vct k})+\omega_m({\vct p}))\chi_{B_{\epsilon}({\vct 0})}({\vct k}+{\vct p})(\omega_m({\vct p})\omega_m({\vct k})-{\vct p}.{\vct k}+2m^2)\, \nonumber \\
    & & \frac{d{\vct k}}{\omega_m({\vct k})}\,\frac{d{\vct p}}{\omega_m({\vct p})} \nonumber \\
 & = & \frac{e^2}{2\pi^2}\int\chi_{(a,b)}(\omega_m({\vct k})+\omega_m({\vct p}))\chi_{B_{\epsilon}({\vct 0})-{\vct p}}({\vct k})(\omega_m({\vct p})\omega_m({\vct k})-{\vct p}.{\vct k}+2m^2)\, \nonumber \\
    & & \frac{d{\vct k}}{\omega_m({\vct k})}\,\frac{d{\vct p}}{\omega_m({\vct p})} \nonumber \\
    & \approx & \frac{e^2}{2\pi^2}\int\chi_{(a,b)}(2\omega_m({\vct p}))(3m^2+2{\vct p}^2)\,\frac{d{\vct p}}{\omega_m({\vct p})^2}(\frac{4}{3}\pi\epsilon^3). \nonumber 
\end{eqnarray}
Therefore defining
\begin{equation}
g_a(b) = \lim_{\epsilon\rightarrow0}\epsilon^{-3}g(a,b,\epsilon),
\end{equation}
we have
\begin{eqnarray}
g_a(b) & = & \frac{e^2}{2\pi^2}\int\chi_{(a,b)}(2\omega_m({\vct p}))(3m^2+2{\vct p}^2)\,\frac{d{\vct p}}{\omega_m({\vct p})^2}(\frac{4}{3}\pi).
\end{eqnarray}
Now
\begin{align*}
\chi_{(a,b)}(2\omega_m({\vct p}))=1&\Leftrightarrow a<2\omega_m({\vct p})<b\\
&\Leftrightarrow (\frac{a}{2})^2<m^2+{\vct p}^2<(\frac{b}{2})^2\\
&\Leftrightarrow m^2(\frac{a^2}{4m^2}-1)<{\vct p}^2<m^2(\frac{b^2}{4m^2}-1)\\
&\Leftrightarrow|{\vct p}|\in(mZ(a),mZ(b)),
\end{align*}
where $Z:[2m,\infty)\rightarrow[0,\infty)$ is given by Eq.~\ref{eq:Z_def}.
Therefore
\begin{equation}
g_a(b)= \frac{2}{\pi}e^2\int_{r=mZ(a)}^{mZ(b)}(3m^2+2r^2)\frac{r^2}{m^2+r^2}\,dr\,(\frac{4}{3}\pi).
\end{equation}
Thus, using the spectral calculus and Leibniz' integral rule, we compute the spectrum of $\Pi$ as follows.
\begin{eqnarray}
\sigma(b) & = & \frac{3}{4\pi}bg_a^{\prime}(b) \nonumber \\
    & = & \frac{2}{\pi}e^2b(3m^2+2m^2Z^2(b))\frac{m^2Z^2(b)}{m^2+m^2Z^2(b)}\frac{b}{4mZ(b)} \nonumber \\
    & = & \frac{2}{\pi}e^2m^3Z(b)(3+2Z^2(b)). \nonumber
\end{eqnarray}

The spectrum has this value $\sigma(b)$ for $b\ge2m$ and the value 0 for $b\leq2m$. Thus, using Eq.~\ref{eq:equiv_density}, $\Pi$ exists as a tempered measure with density
\begin{equation} \label{eq:Pi_density}
\Pi(q)=\left\{\begin{array}{l}
(q^2)^{-\frac{1}{2}}\sigma((q^2)^{\frac{1}{2}})\mbox{ if }q^2>0,q^0>0 \\
0\mbox{ otherwise},
\end{array}\right.
\end{equation}
where 
\begin{equation} \label{eq:pi_spectrum}
\sigma(b)=\left\{\begin{array}{l}\frac{2}{\pi}e^2m^3Z(b)(3+2Z^2(b))\mbox{ for } b\ge2m \\
0\mbox{ otherwise,}\end{array}\right.
\end{equation} 
is the spectrum of the measure $\Pi$.

One can now see that $\Pi$ is a (Borel) measure in the ordinary sense of the term, i.e. a $[0,\infty]$ valued countably additive function on ${\mathcal B}({\bf R}^4)$ which vanishes on the empty set and is finite on compact sets. (It is clearly defined on the larger sigma algebra of Lebesgue measurable sets.) $\Pi$ is finite on compact sets and when evaluated on test functions of rapid decrease, it is not divergent.

\subsection{The vacuum polarization function $\pi$}

$q\mapsto\Pi(q)$ does not exist pointwise as a function, the integral defining it is divergent. However, pretend for the moment that $\Pi$ did exist as a function. Then we would be able to define a measure which we also denote by $\Pi$ by
\begin{equation}
\Pi(\Gamma)=\int_{\Gamma}\Pi(q)\,dq.
\end{equation}
Thus the function $\Pi$ is the density defining the measure $\Pi$. 

But we have just shown how $\Pi$ can be considered to be a measure with density given by Eq.~\ref{eq:Pi_density}. Thus we may think of $\Pi$ the function as being defined to be equal to this density.

It is straightforward to show that the  vacuum polarization tensor valued measure $\Pi^{\mu\nu}$ is causal and Lorentz covariant. 

\begin{theorem}
The measure associated with the vacuum polarization tensor has the form
\[ \Pi^{\mu\nu}(\Gamma)=\int_{m^{\prime}=0}^{\infty}\int_{{\bf R}^4}\chi_{\Gamma}(p)(p^2\eta^{\mu\nu}-p^{\mu}p^{\nu})\,\Omega_{m^{\prime}}(dp)\,\sigma_1(m^{\prime})\,dm^{\prime}, \]
for some continuous spectral function $\sigma_1:(0,\infty)\rightarrow{\bf C}$.
\end{theorem}
{\bf Proof}

One can readily compute, using the gamma matrix trace identities, that
\[ \mbox{Tr}(\gamma^{\mu}({\slas p}+m)\gamma^{\nu}({\slas k}-m))=4(p^{\mu}k^{\nu}-\eta^{\mu\nu}p.k+k^{\mu}p^{\nu}-m^2\eta^{\mu\nu}). \]
Thus
\begin{align*}
g^{\mu\nu}(a,b,\epsilon)&=\Pi^{\mu\nu}(\Gamma(a,b,\epsilon))\\
&=-\frac{e^2}{16\pi^2}\int\chi_{\Gamma(a,b,\epsilon)}(k+p)\mbox{Tr}(\gamma^{\mu}({\slas p}+m)\gamma^{\nu}({\slas k}-m))\,\Omega_m(dk)\,\Omega_m(dp)\\
&=-\frac{e^2}{4\pi^2}\int\chi_{\Gamma(a,b,\epsilon)}(k+p)(p^{\mu}k^{\nu}-\eta^{\mu\nu}p.k+k^{\mu}p^{\nu}-m^2\eta^{\mu\nu} )\,\Omega_m(dk)\,\Omega_m(dp)\\
&\approx-\frac{e^2}{4\pi^2}\int\chi_{(a,b)}(\omega_m({\vct k})+\omega_m({\vct p}))\chi_{B_{\epsilon}({\vct 0})}({\vct k}+{\vct p})((\omega_m({\vct p}),{\vct p})^{\mu}(\omega_m({\vct k}),{\vct k})^{\nu}-\\
&\eta^{\mu\nu}(\omega_m({\vct p})\omega_m({\vct k})-{\vct p}.{\vct k})+(\omega_m({\vct k}),{\vct k})^{\mu}(\omega_m({\vct p}),{\vct p})^{\nu}-m^2\eta^{\mu\nu})\,\frac{d{\vct k}}{\omega_m({\vct k})}\frac{d{\vct p}}{\omega_m({\vct p})}\\
&=-\frac{e^2}{4\pi^2}\int\chi_{(a,b)}(\omega_m({\vct k})+\omega_m({\vct p}))\chi_{B_{\epsilon}({\vct 0})-{\vct p}}({\vct k})((\omega_m({\vct p}),{\vct p})^{\mu}(\omega_m({\vct k}),{\vct k})^{\nu}-\\
&\eta^{\mu\nu}(\omega_m({\vct p})\omega_m({\vct k})-{\vct p}.{\vct k})+(\omega_m({\vct k}),{\vct k})^{\mu}(\omega_m({\vct p}),{\vct p})^{\nu}-m^2\eta^{\mu\nu})\,\frac{d{\vct k}}{\omega_m({\vct k})}\frac{d{\vct p}}{\omega_m({\vct p})}\\
&\approx-\frac{e^2}{4\pi^2}\int\chi_{(a,b)}(2\omega_m({\vct p}))((\omega_m({\vct p}),{\vct p})^{\mu}(\omega_m({\vct p}),-{\vct p})^{\nu}-\eta^{\mu\nu}(\omega_m({\vct p})^2+{\vct p}^2)+\\
&(\omega_m({\vct p}),-{\vct p})^{\mu}(\omega_m({\vct p}),{\vct p})^{\nu}-m^2\eta^{\mu\nu})\frac{1}{\omega_m({\vct p})^2}\,d{\vct p}\,(\frac{4}{3}\pi\epsilon^3).
\end{align*}
Therefore
\begin{align*}
g_a^{\mu\nu}(b)&=\lim_{\epsilon\rightarrow0}\epsilon^{-3}g^{\mu\nu}(a,b,\epsilon)\\
&=-\frac{e^2}{4\pi^2}\int_{r=mZ(a)}^{mZ(b)}\int_{\theta=0}^{\pi}\int_{\phi=0}^{2\pi}((\omega_m({\vct p}),{\vct p})^{\mu}(\omega_m({\vct p}),-{\vct p})^{\nu}-\eta^{\mu\nu}(m^2+2r^2)+\\
&(\omega_m({\vct p}),-{\vct p})^{\mu}(\omega_m({\vct p}),{\vct p})^{\nu}-m^2\eta^{\mu\nu})\frac{r^2}{m^2+r^2}\sin(\theta)\,d\phi\,d\theta\,dr\,(\frac{4}{3}\pi),
\end{align*}
where ${\vct p}={\vct p}(r,\theta,\phi)$ (spherical polar coordinates).
Hence
\begin{align*}
g_a^{\mu\mu\prime}(b)&=-\frac{e^2}{4\pi^2}\int_{\theta=0}^{\pi}\int_{\phi=0}^{2\pi}((\omega_m({\vct p}),{\vct p})^{\mu}(\omega_m({\vct p}),-{\vct p})^{\mu}-\eta^{\mu\mu}(m^2+2r^2)+\\
&(\omega_m({\vct p}),-{\vct p})^{\mu}(\omega_m({\vct p}),{\vct p})^{\mu}-m^2\eta^{\mu\mu})\frac{r^2}{m^2+r^2}\sin(\theta)\,d\phi\,d\theta\,(\frac{4}{3}\pi)\\
&mZ^{\prime}(b),
\end{align*}
where $r=mZ(b)$. Thus
\begin{align*}
g_a^{00\prime}(b)&=-\frac{e^2}{4\pi^2}\int_{\theta=0}^{\pi}\int_{\phi=0}^{2\pi}(m^2+r^2-(m^2+2r^2)+m^2+r^2-m^2)\\
&\frac{r^2}{m^2+r^2}\sin(\theta)\,d\phi\,d\theta\,(\frac{4}{3}\pi)(mZ^{\prime}(b))\\
&=0.
\end{align*}
Thus $g_a^{\mu\nu}$ is continuously differentiable and $g_a^{00\prime}\equiv0$. Therefore by the spectral theorem for tensor valued measures and Eq.~\ref{eq:tensor_2},
$\Pi^{\mu\nu}$ has the form of Eq.~\ref{eq:general_L_i_measure}, i.e.
\begin{equation}
\Pi^{\alpha\beta}(\Gamma)=\int_{m^{\prime}=0}^{\infty}\int_{\Gamma}\eta^{\alpha\beta}p^2\,\Omega_{m^{\prime}}(dp)\,\sigma_1(dm^{\prime})+\int_{m^{\prime}=0}^{\infty}\int_{\Gamma}p^{\alpha}p^{\beta}\,\Omega_{m^{\prime}}(dp)\,\sigma_2(dm^{\prime}),
\end{equation}
 with $\sigma_1$ and $\sigma_2$ continuous functions and 
\begin{equation}
\sigma_2(m^{\prime})=-\sigma_1(m^{\prime}),\forall b>0.
\end{equation}
$\Box$

Therefore
\begin{align*}
\Pi^{\mu\nu}(\Gamma)&=\int_{m^{\prime}=0}^{\infty}\int_{{\bf R}^3}\chi_{\Gamma}(\omega_{m^{\prime}}({\vct p}),{\vct p})(m^{\prime2}\eta^{\mu\nu}-(\omega_{m^{\prime}}({\vct p}),{\vct p})^{\mu}(\omega_{m^{\prime}}({\vct p}),{\vct p})^{\nu})\sigma_1(m^{\prime})\\
&\frac{d{\vct p}}{\omega_{m^{\prime}}({\vct p})}\,dm^{\prime}.\\
\end{align*}
Now make the coordinate transformation 
\begin{equation}
q=q(m^{\prime},{\vct p})=(\omega_{m^{\prime}}({\vct p}),{\vct p}),m^{\prime}>0,{\vct p}\in{\bf R}^3.
\end{equation}
The Jacobian of the transformation is
\begin{equation}
J(m^{\prime},{\vct p})=m^{\prime}\omega_{m^{\prime}}({\vct p})^{-1}.
\end{equation}
Thus
\begin{align*}
\Pi^{\mu\nu}(\Gamma)&=\int_{q^2>0,q^0>0}\chi_{\Gamma}(q)(q^2\eta^{\mu\nu}-q^{\mu}q^{\nu})(q^2)^{-\frac{1}{2}}\sigma_1((q^2)^{\frac{1}{2}})\,dq.
\end{align*}
Hence the density $\Pi^{\mu\nu}$ for the measure $\Pi^{\mu\nu}$ is given by
\begin{equation} \label{eq:Pi_form}
\Pi^{\mu\nu}(q)=(q^2\eta^{\mu\nu}-q^{\mu}q^{\nu})\pi(q),
\end{equation}
where
\begin{equation} 
\pi(q)=\left\{
\begin{array}{l}
(q^2)^{-\frac{1}{2}}\sigma_1((q^2)^{\frac{1}{2}})\mbox{ if }q^2>0,q^0>0\\
0\mbox{ otherwise.}
\end{array}\right. 
\end{equation}
(The fact that $\Pi^{\mu\nu}$ has the form of Eq.~\ref{eq:Pi_form} is well known but has previously been established through manipulating infinite quantities during renormalization (Weinberg, 2005, p. 478).) 

Contracting Eq.~\ref{eq:Pi_form} with the Minkowski metric tensor leads to
\begin{equation}
\pi(q)=\frac{1}{3q^2}\Pi(q).
\end{equation}
Therefore, using Eq.~\ref{eq:Pi_density}, our spectral vacuum polarization function, defined on the domain
\begin{equation}
D=\{q\in{\bf R}^4:q^2>0,q^0>0\},
\end{equation}
is
\begin{equation} \label{eq:pi_def}
\pi(q)=\frac{1}{3q^2}\Pi(q)=\frac{1}{3}(q^2)^{-\frac{3}{2}}\sigma((q^2)^{\frac{1}{2}}).
\end{equation}
$\pi$ is a function on $D$ supported on $C_{2m}$ but its value for argument $q$ only depends on $q^2$. Therefore, with no fear of confusion, one may define the vacuum polarization function $\pi:(0,\infty)\rightarrow[0,\infty)$ by
\begin{equation} \label{eq:polarization_function}
\pi(s)=\frac{1}{3}s^{-3}\sigma(s)=\left\{\begin{array}{l}
\frac{2}{3\pi}s^{-3}e^2m^3Z(s)(3+2Z^2(s))\mbox{ if }s\geq2m\\
0\mbox{ otherwise.}
\end{array}\right.
\end{equation}
Define $\zeta:{\bf R}^4\rightarrow{\bf R}$ by
\begin{equation}
\zeta(q)=\mbox{sign}(q^2)|q^2|^{\frac{1}{2}},
\end{equation}
where sign is the function defined by
\begin{equation}
\mbox{sign}(s)=\left\{\begin{array}{ccc}
1&\mbox{ if }&s>0\\
0&\mbox{ if }&s=0\\
-1&\mbox{ if }&s<0.
\end{array}\right.
\end{equation}
$\zeta$ is constant on orbits of the action of $O(1,3)$ on Minkowski space. Let  $C=\{p\in{\bf R}^4:p^2=0,p\neq0\}$ be the (punctured) null cone. Then $s\mapsto\zeta^{-1}(s)=\{p\in{\bf R}^4:\zeta(p)=s\}$ is a bijection of ${\bf R}\backslash\{0\}$ onto the set $({\bf R}^4/O(1,3))\backslash\{\{0\},C\}\}$ of hyperboloid orbits ($s$ is positive for timelike orbits and negative for spacelike orbits).

Then the vacuum polarization function on $D$ is given by
\begin{equation} \label{eq:vp_fn_def}
\pi(q)=\pi(\zeta(q)), 
\end{equation}
where the $\pi$ on the left hand side of this equation is the vacuum polarization function that we have defined by Eq.~\ref{eq:pi_def} on the subset $D$ of Minkowski space while the $\pi$ on the right hand side of this equation is the vacuum polarization function on $(0,\infty)$ we have defined by Eq.~\ref{eq:polarization_function}. We may take the expression given in Eq.~\ref{eq:vp_fn_def} to define $\pi$ over all of ${\bf R}^4\backslash(C\cup\{0\})$. Note that $s\mapsto\pi(s)$ is an odd function on ${\bf R}\backslash\{0\}$.

\subsection{Comparison of the spectral vacuum polarization function with the renormalized vacuum polarization function}

Regularization and renormalization are techniques invented by physicists to control the infinities in divergent integrals in QFT to obtain finite answers which can be compared with experiment. The answers obtained using these methods are in close agreement with experiment so there is clearly great merit in the approach. However many mathematicians are confused by these methods since they do not seem to make mathematical sense (e.g. introducing infinite ``counterterms" into Lagrangians to cancel infinities produced when carrying out integrations implied by these Lagrangians or perturbing the dimension $D$ of space-time to $D=4-\epsilon, \epsilon>0$ because everything blows up when $D=4$ and then later ignoring or subtracting out terms proportional to $\epsilon^{-1}$ before taking the limit as $\epsilon$ tends to $0$ to obtain the answers which are compared with experiment (dimensional regularization / renormalization)). 

The vacuum polarization function is generally computed in QFT using the dimensional regularization / renormalization approach with the result
\begin{equation}
\pi_r(k^2)=-\frac{2\alpha}{\pi}\int_0^1dz\,z(1-z)\log(1-\frac{k^2z(1-z)}{m^2}),
\end{equation}
(Mandl and Shaw, 1991, p. 229) where (in natural units) $\alpha=(4\pi)^{-1}e^2$ is the fine structure constant. $\pi_r$ is defined for all $k\in{\bf R}^4$ for which $k^2<4m^2$. The integral can be performed leading to the analytic expression
\begin{equation} 
\pi_r(k^2)=-\frac{\alpha}{3\pi}\{\frac{1}{3}+2(1+\frac{2m^2}{k^2})[(\frac{4m^2}{k^2}-1)^{\frac{1}{2}}\mbox{arcot}(\frac{4m^2}{k^2}-1)^{\frac{1}{2}}-1]\}.
\end{equation} 
(See Appendix 1 for a proof of this.)

Thus 
\begin{equation}
\pi_r(k)=-\frac{\alpha}{3\pi}(\frac{1}{3}+(Y^2+3)(Y\mbox{arcot}(Y)-1)),
\end{equation}
where
\begin{equation}
Y=Y(k)=\left(\frac{4m^2}{k^2}-1\right)^{\frac{1}{2}}.
\end{equation}
This expression for $\pi_r$ is defined on $\{k\in{\bf R}^4:0<k^2\leq4m^2\}$.

$\pi_r(k)$ depends only on the value $k^2$ of its argument $k$. Therefore we define (without fear of confusion) $\pi_r:(0,2m]\rightarrow[0,\infty)$ by
\begin{equation}
\pi_r(s)=-\frac{\alpha}{3\pi}(\frac{1}{3}+(Y^2+3)(Y\mbox{arcot}(Y)-1)),
\end{equation}
where
\begin{equation}
Y=Y(s)=\left(\frac{4m^2}{s^2}-1\right)^{\frac{1}{2}}.
\end{equation}
Let $\pi_s$ denote our $\pi$ calculated using spectral calculus. $\pi_s(s)$ is defined for $s\geq2m$ while $\pi_r(s)$ is defined for $0<s\leq2m$. To compare them we note that
\begin{equation}
Z(s)=\left(\frac{s^2}{4m^2}-1\right)^{\frac{1}{2}}=\frac{s}{2m}\left(1-\frac{4m^2}{s^2}\right)^{\frac{1}{2}}=\frac{si}{2m}\left(\frac{4m^2}{s^2}-1\right)^{\frac{1}{2}}=\frac{si}{2m}Y(s),
\end{equation}
where $Z:[2m,\infty)\rightarrow[0,\infty)$ is the function defined by Eq.~\ref{eq:Z_def}.

Both Weinberg (2005, p. 475)  and Itzykson and Zuber (1980, p. 322) in their highly complex manipulations to compute $\pi_r$ do a rotation in the complex plane. Thus we will compare our $\pi_s$ with their $\pi_r$ using the assignment
\begin{equation}
iY\mapsto Y.
\end{equation}

$\pi_r$ is obtained by regularization and renormalization involving manipulating and subtracting both infinite and finite quantities. We find that $\pi_r$ and $\pi_s$ coincide approximately when $\pi_r(s)$ is rescaled by a factor ($\lambda$ say) and then shifted by an amount ($\tau$ say). It is straightforward to prove following. 
\begin{lemma}
Suppose that 
\begin{equation}
{\ch\pi}_r(\rho)=\tau+\lambda\pi_r(\rho),
\end{equation}
where $\tau,\lambda\in{\bf R}$ (and $\rho=s/m$). Then 
\begin{equation}
{\ch\pi}_r(2)=\pi_s(2),
\end{equation}
and
\begin{equation}
\lim_{\rho\rightarrow\infty}{\ch\pi}_r(\rho)=\lim_{\rho\rightarrow\infty}\pi_s(\rho),
\end{equation}
if and only if
\begin{equation}
\tau=-\frac{2e^2}{9\pi^2}\lambda \mbox{ with } \lambda=-\frac{2\pi}{\pi-1}.
\end{equation}
\end{lemma}

It is to be emphasized that these normalization constants are finite and, furthermore, we only use them for the purpose of comparing $\pi_s$ with $\pi_r$ but not in any of the calculations that we carry out with $\pi_s$. No constants need to be fixed to compute our vacuum polarization function $\pi_s$, no form of renormalization or normalization is required or used in any of the calculations that we make with it.  

C++ code for a program to compare our spectral vacuum polarization function with the vacuum polarization function computed by dimensional regularization / renormalization is as follows. 

\begin{verbatim}
#include <iostream>
#include <fstream>
#include <math.h>

const int N_display = 10000;
const double Lambda = 10.0;
const double pi = 4.0 * atan(1.0);

int main()
{
    std::cout << "Hello World!\n";
    std::ofstream outFile("out.txt");
    double delta = Lambda / N_display;
    int i;
    for (i = 1; i <= N_display; i++)
    {
        double rho = 2.0 + i * delta; // q/m
        double Z = sqrt(rho * rho / 4.0 - 1.0);
        double v = 1.0 / (rho * rho * rho);
        double pi_spectral;
        pi_spectral = v * Z * (3.0 + 2.0 * Z * Z);
        pi_spectral *= (2.0 / (3.0 * pi));
        double X = 2.0 * Z / rho;
        double pi_renormalized;
        pi_renormalized = (1.0 / 3.0) + (X * X + 3.0) * (X * atan(1.0 / X) - 1.0);
        pi_renormalized /= -12.0 * pi * pi;
        double lambda = -2.0 * pi / (pi - 1.0);
        double tau = -2.0 * lambda / (9.0 * pi * pi);
        double pi_tilde = tau + lambda * pi_renormalized;
        outFile << rho << '\t' << pi_tilde << '\t'
            << pi_spectral << "\n";
    }
    return(0);
}
\end{verbatim}

\begin{figure} 
\centering
\includegraphics[width=15cm]{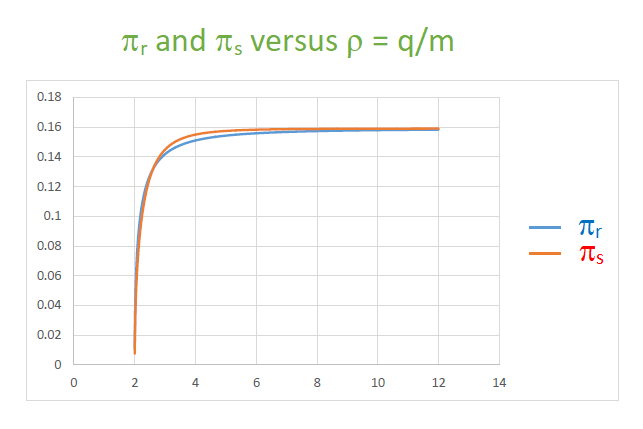}
\caption{Renormalized and spectral vacuum polarization versus $\rho = q/m$.} \label{fig:Fig1}
\end{figure}

The graph of the output produced by this program (with vertical axis rescaled) is given in Figure~\ref{fig:Fig1}. The average deviation between the spectral vacuum polarization $\pi_s$ and the vacuum polarization $\pi_r$ obtained using dimensional regularization / renormalization is 1.48\%  of the average value of $\pi_r$ over the range shown in Figure~\ref{fig:Fig1} and the limit as the range $\rightarrow\infty$ of this average is 0. Thus we have shown that
\begin{equation}
\pi_s(\rho)\approx\tau+\lambda\pi_r(\rho), \forall\rho\in[2,\infty).
\end{equation}
It can be seen that (up to finite normalization) the difference between the spectral vacuum polarization and that obtained using dimensional regularization / renormalization is very small even though they are defined by completely different analytic expressions and derived by totally different approaches.

\section{The Uehling contribution to the Lamb shift for the H atom \label{section:Uehling}}

We will now check the validity of our spectral vacuum polarization function by computing the Uehling contribution to the Lamb shift for the H atom. Such investigations serve to further confirm the validity of QED and/or help to search for physics beyond the standard model.  Following Weinberg (2005) we carry out a {\em gedankenexperiment} in which an electron is scattered off a proton and compute, using the Born approximation, the Uehling effect, i.e. the result of including a Feynman diagram with a single fermion loop in addition to the tree level diagram, to compute the effective potential of the H atom for determination of the Uehling contribution to the Lamb shift. 

The tree level diagram associated with electron-proton scattering contributes a scattering matrix of
\begin{equation}
S_{afi}=(2\pi)^4\delta(p_1^{\prime}+p_2^{\prime}-p_1-p_2){\mathcal M}_a,
\end{equation}
where ${\mathcal M}_a$ is given by
\begin{align} \label{eq:M_a_def}
i{\mathcal M}_{a,\alpha_1^{\prime}\alpha_2^{\prime}\alpha_1\alpha_2}(p_1^{\prime},p_2^{\prime},p_1,p_2)=&\overline{u}_1( p_1^{\prime},\alpha_1^{\prime})ie_1\gamma^{\rho}u_1(p_1,\alpha_1)iD_{\rho\sigma}(q)\\
&\overline{u}_2(p_2^{\prime},\alpha_2^{\prime})ie_2\gamma^{\sigma} u_2(p_2,\alpha_2),\nonumber 
\end{align} 
where
\begin{equation}
D_{\rho\sigma}(q)=-\frac{\eta_{\rho\sigma}}{q^2+i\epsilon},
\end{equation}
is the photon propagator, $e_1=-e$ and $e_2=e$ are the charges of the electron and proton respectively, $u_1(p,\alpha),u_2(p,\alpha)$ are Dirac spinors corresponding to $p$ in $H_{m_e}^{+}$ and $H_{m_p}^{+}$, the mass shells of the electron and the proton respectively, for $\alpha\in\{0,1\}$ (see Appendix 3), and $q=p_1^{\prime}-p_1=-(p_2^{\prime}-p_2)$ is the momentum transfer. Therefore
\begin{equation}
{\mathcal M}_{a,\alpha_1^{\prime}\alpha_2^{\prime}\alpha_1\alpha_2}(p_1^{\prime},p_2^{\prime},p_1,p_2)=\frac{{\mathcal M}_{0,\alpha_1^{\prime}\alpha_2^{\prime}\alpha_1\alpha_2}(p_1^{\prime},p_2^{\prime},p_1,p_2)}{q^2+i\epsilon},
\end{equation}
where
\begin{equation} \label{eq:M_0_def}
{\mathcal M}_{0,\alpha_1^{\prime}\alpha_2^{\prime}\alpha_1\alpha_2}(p_1^{\prime},p_2^{\prime},p_1,p_2)  =  e_1e_2\overline{u}_1( p_1^{\prime},\alpha_1^{\prime})\gamma^{\rho}u_1(p_1,\alpha_1)\eta_{\rho\sigma}\overline{u}_2(p_2^{\prime},\alpha_2^{\prime})\gamma^{\sigma} u_2(p_2,\alpha_2). 
\end{equation}
(There is another contributing diagram obtained by making the substitution $p_1^{\prime}\leftrightarrow p_2^{\prime}$).

The vacuum polarization (fermion loop) contribution to the scattering matrix is given by
\begin{equation}
S_{bfi}=(2\pi)^4\delta(p_1^{\prime}+p_2^{\prime}-p_1-p_2){\mathcal M}_b,
\end{equation}
where ${\mathcal M}_b$ is given by
\begin{eqnarray} \label{eq:Uehling_amplitude}
i{\mathcal M}_{b,\alpha_1^{\prime}\alpha_2^{\prime}\alpha_1\alpha_2}(p_1^{\prime},p_2^{\prime},p_1,p_2) & = & \overline{u}_1(p_1^{\prime},\alpha_1^{\prime})ie_1\gamma^\rho u_1(p_1,\alpha_1) \\
    & & iD_{F\rho\mu}(q)i\Pi^{\mu\nu}(q)iD_{F\nu\sigma}(q) \nonumber \\          
    &  & \overline{u}_2(p_2^{\prime},\alpha_2^{\prime})ie_2\gamma^{\sigma}u_2(p_2,\alpha_2). \nonumber 
\end{eqnarray}
Therefore, since
\begin{align*}
D_{F\rho\mu}(q)\Pi^{\mu\nu}(q)D_{F\nu\sigma}(q)&=(\frac{1}{q^2})^2\eta_{\rho\mu}(q^2\eta^{\mu\nu}-q^{\mu}q^{\nu})\pi(q)\eta_{\nu\sigma}\\
&=(\frac{1}{q^2}\eta_{\rho\sigma}-(\frac{1}{q^2})^2q_{\rho}q_{\sigma})\pi(q),
\end{align*}
and, by a well known conservation property,
\[ \overline{u}_1(p_1^{\prime},\alpha_1^{\prime})q_{\rho}\gamma^{\rho}u_1(p_1,\alpha_1)=0, \]
we have
\begin{equation}
{\mathcal M}_{b,\alpha_1^{\prime}\alpha_2^{\prime}\alpha_1\alpha_2}(p_1^{\prime},p_2^{\prime},p_1,p_2)=\frac{{{\mathcal M}_{0,\alpha_1^{\prime}\alpha_2^{\prime}\alpha_1\alpha_2}}(p_1^{\prime},p_2^{\prime},p_1,p_2)\pi(q)}{q^2+i\epsilon}.
\end{equation}

The total scattering matrix is given by
\begin{equation}
S_{a+bfi}=S_{afi}+S_{bfi}=(2\pi)^4\delta(p_1^{\prime}+p_2^{\prime}-p_1-p_2){\mathcal M}_{a+b},
\end{equation}
where
\begin{eqnarray}
{\mathcal M}_{a+b,\alpha_1^{\prime}\alpha_2^{\prime}\alpha_1\alpha_2}(p_1^{\prime},p_2^{\prime},p_1,p_2) & = &{\mathcal M}_{0,\alpha_1^{\prime}\alpha_2^{\prime}\alpha_1\alpha_2}(p_1^{\prime},p_2^{\prime},p_1,p_2)\frac{1+\pi(q)}{q^2+i\epsilon}. \label{eq:Feynman_amplitude}
\end{eqnarray}

We will now consider the non-relativistic (NR) approximation. In this approximation we have
\begin{equation}
{\mathcal M}_{0,\alpha_1^{\prime}\alpha_2^{\prime}\alpha_1\alpha_2}(p_1^{\prime},p_2^{\prime},p_1,p_2)=e_1e_2\delta_{\alpha_1^{\prime}\alpha_1}\delta_{\alpha_2^{\prime}\alpha_2},
\end{equation}
(Appendix 3).
Therefore
\begin{equation}
{\mathcal M}_{a+b,\alpha_1^{\prime}\alpha_2^{\prime}\alpha_1\alpha_2}(p_1^{\prime},p_2^{\prime},p_1,p_2)=e_1e_2\frac{1+\pi(q)}{q^2+i\epsilon}\delta_{\alpha_1^{\prime}\alpha_1}\delta_{\alpha_2^{\prime}\alpha_2}.
\end{equation}
Another aspect of the NR approximation is that $q^0$ is negligible compared with $|{\vct q}|$ (Weinberg, 2005, p. 480). Therefore the scattering matrix associated with the process is
\begin{align*}
S_{a+bfi}&=(2\pi)^4\delta(p_1^{\prime}+p_2^{\prime}-p_1-p_2){\mathcal M}_{a+b,\alpha_1^{\prime}\alpha_2^{\prime}\alpha_1\alpha_2}(p_1^{\prime},p_2^{\prime},p_1,p_2)\\
&=(2\pi)^4e_1e_2\delta(p_1^{\prime}+p_2^{\prime}-p_1-p_2)\frac{1+\pi(q)}{q^2+i\epsilon}\delta_{\alpha_1^{\prime}\alpha_1}\delta_{\alpha_2^{\prime}\alpha_2}\\
&=-(2\pi)^4e_1e_2\delta(p_1^{\prime}+p_2^{\prime}-p_1-p_2)\frac{1+\pi(0,{\vct q})}{{\vct q}^2}\delta_{\alpha_1^{\prime}\alpha_1}\delta_{\alpha_2^{\prime}\alpha_2}.
\end{align*}

The Born approximation enables one, in NR quantum mechanics, to compute the scattering matrix for a scattering process as being proportional to the Fourier transform of the potential between the scattering particles. In NR scattering the polarizations of the scattering particles are preserved. Therefore the scattering matrix is
\begin{align} 
S_{Born}&=S_{Born}(p_1^{\prime},p_2^{\prime},p_1,p_2)\nonumber\\
&=\omega\delta(p_1^{\prime}+p_2^{\prime}-p_1-p_2)\int V({\vct r})e^{-i{\vct q}.{\vct r}}\,d{\vct r}\delta_{\alpha_1^{\prime}\alpha_1}\delta_{\alpha_2^{\prime}\alpha_2}, \label{eq:Born1}
\end{align}
where $\omega\in{\bf C}$ is some constant. 

To determine the constant $\omega$ we consider tree level electron-proton scattering with scattering matrix
\begin{equation}
S_{afi}= -(2\pi)^4e_1e_2\delta(p_1^{\prime}+p_2^{\prime}-p_1-p_2)\frac{1}{{\vct q}^2}\delta_{\alpha_1^{\prime}\alpha_1}\delta_{\alpha_2^{\prime}\alpha_2},
\end{equation}
and equate this to $S_{Born}$ resulting in the equation
\begin{equation}
-(2\pi)^4e_1e_2\frac{1}{{\vct q}^2}=\omega\int V_a({\vct r})e^{-i{\vct q}.{\vct r}}\,d{\vct r}.
\end{equation}
If the inverse Fourier transform of this function exists then we have, by taking the inverse Fourier transform of both sides, that
\begin{equation}
-(2\pi)e_1e_2\int\frac{1}{{\vct q}^2}e^{i{\vct q}.{\vct r}}\,d{\vct q}=\omega V_a({\vct r}).
\end{equation}
But, as is well known, we have
\begin{equation}
\int\frac{1}{{\vct q}^2}e^{i{\vct q}.{\vct r}}\,d{\vct q}=\frac{2\pi^2}{|{\vct r}|},
\end{equation}
as an improper Riemann multiple integral (using polar coordinates), (but not as a Lebesgue integral).
Therefore
\begin{equation}
-4\pi^3\frac{e_1e_2}{|{\vct r}|}=\omega V_a({\vct r}),
\end{equation}
i.e.
\begin{equation}
V_a({\vct r})=-\omega^{-1}4\pi^3\frac{e_1e_2}{|{\vct r}|}.
\end{equation}
This is the well known and celebrated derivation of the form of Coulomb's law from QFT. To determine the constant we set
\begin{equation}
V_a({\vct r})=\frac{e_1e_2}{4\pi}\frac{1}{|{\vct r}|},
\end{equation}
to be the Coulomb potential (this amounts to a choice for units for charge, in this case rationalized units) and we have that
\begin{equation}
-\omega^{-1}4\pi^3\frac{e_1e_2}{|{\vct r}|}=\frac{e_1e_2}{4\pi}\frac{1}{|{\vct r}|},
\end{equation}
which is satisfied if and only if $\omega=-(2\pi)^4$.
Thus from Eq.~\ref{eq:Born1} we have that, in general,
\begin{equation} \label{eq:Born2}
S_{Born}(p_1^{\prime},p_2^{\prime},p_1,p_2)=-(2\pi)^4\delta(p_1^{\prime}+p_2^{\prime}-p_1-p_2)\int V({\vct r})e^{-i{\vct q}.{\vct r}}\,d{\vct r}\delta_{\alpha_1^{\prime}\alpha_1}\delta_{\alpha_2^{\prime}\alpha_2}.
\end{equation}

Now for the case of the process with scattering matrix $S_{a+bfi}$ we have, using the Born approximation, that 
\begin{equation}
S_{Born,a+bfi}=S_{a+bfi},
\end{equation}
i.e.
\begin{align*}
&-(2\pi)^4\delta(p_1^{\prime}+p_2^{\prime}-p_1-p_2)\int V_{a+b}({\vct r})e^{-i{\vct q}.{\vct r}}\,d{\vct r}\,\delta_{\alpha_1^{\prime}\alpha_1}\delta_{\alpha_2^{\prime}\alpha_2}\\
&=-(2\pi)^4e_1e_2\delta(p_1^{\prime}+p_2^{\prime}-p_1-p_2)\frac{1+\pi(0,{\vct q})}{{\vct q}^2}\delta_{\alpha_1^{\prime}\alpha_1}\delta_{\alpha_2^{\prime}\alpha_2},
\end{align*}
and therefore
\begin{equation}
\int V_{a+b}({\vct r})e^{-i{\vct q}.{\vct r}}\,d{\vct r}=e_1e_2\frac{1+\pi(0,{\vct q})}{{\vct q}^2}.
\end{equation}
Taking the inverse Fourier transform of both sides of this equation we have, if the inverse Fourier transform exists, 
\begin{align}
V_{a+b}({\vct r})&=(2\pi)^{-3}e_1e_2\int\frac{1+\pi(0,{\vct q})}{{\vct q}^2}e^{i{\vct q}.{\vct r}}\,d{\vct q}\nonumber\\
&=-(2\pi)^{-3}e^2\frac{2\pi^2}{|{\vct r}|}+(\Delta V)({\vct r}),\nonumber\\
&=-\frac{e^2}{4\pi|{\vct r}|}+(\Delta V)({\vct r}), \label{eq:total_energy}
\end{align}
where
\begin{equation} \label{eq:FT_eqn}
(\Delta V)({\vct r})=-(2\pi)^{-3}e^2\int\frac{\pi(0,{\vct q})}{{\vct q}^2}e^{i{\vct q}.{\vct r}}\,d{\vct q}.
\end{equation}
Now $\pi$ is rotationally invariant. Therefore the inverse Fourier transform in Eq.~\ref{eq:FT_eqn} is rotationally invariant. Therefore we can write
\begin{align}
(\Delta V)(r)  &=- (2\pi)^{-3}e^2\int\frac{\pi(0,{\vct q})}{{\vct q}^2}e^{i{\vct q}.(0,0,r)}\,d{\vct q} \nonumber \\
    &= (2\pi)^{-3}e^2\int_{s=0}^{\infty}\int_{\theta=0}^{\pi}\frac{\pi(s)}{s^2}e^{isr\cos(\theta)}2\pi s^2\sin(\theta)\,d\theta\,ds \nonumber \\
    &=-(2\pi)^{-2}ie^2\frac{1}{r}\int_{s=0}^{\infty}\frac{\pi(s)}{s}(e^{isr}-e^{-isr})\,ds, \label{eq:Delta_V_0}
\end{align}
(where we have used the fact that $\pi((0,{\vct q}))=\pi(\zeta((0,{\vct q})))=\pi(-|{\vct q}|)=-\pi(|{\vct q})|$). Thus 
\begin{equation}
(\Delta V)(r)=-(2\pi)^{-2}ie^2\frac{1}{r}\int_{s=0}^{\infty}\frac{\pi(s)}{s}e^{isr}\,ds+(2\pi)^{-2}ie^2\frac{1}{r}\int_{s=0}^{\infty}\frac{\pi(s)}{s}e^{-isr}\,ds, \label{eq:Delta_V_1}
\end{equation}
and
\begin{equation} 
(\Delta V)(r)=(2\pi)^{-2}e^2\frac{1}{r}\int_{s=0}^{\infty}\frac{\pi(s)}{s}\sin(rs)\,ds. \label{eq:Delta_V}
\end{equation}
We will investigate the convergence of the integral given by Eq.~\ref{eq:Delta_V} for the two forms of the vacuum polarization function under consideration in the next section. 

We would like to analytically continue $\Delta V$ to the upper imaginary axis of the complex plane since we interested in spacelike points $x$ in Minkowski space for which $x^2<0$, corresponding to pure imaginary $r$. Therefore we seek a complex analytic function $\Delta V_{\mbox{analytic}}$ associated with $\Delta V$. By a well known Paley-Wiener theorem, since, in the case $\pi=\pi_s$,
\[(s\mapsto\frac{\pi(s)}{s})\in L^2((0,\infty)),\]
the function
\[ r\mapsto\int_{s=0}^{\infty}\frac{\pi(s)}{s}e^{irs}\,ds, \]
is analytic in the upper half plane while the function
\[ r\mapsto\int_{s=0}^{\infty}\frac{\pi(s)}{s}e^{-irs}, \]
is analytic in the lower half plane (but not in the upper half plane). Therefore we take $\Delta V_{\mbox{analytic}}$ to be the function
\begin{equation}
\Delta V_{\mbox{analytic}}(r)=-(2\pi)^{-2}ie^2\frac{1}{r}\int_{s=0}^{\infty}\frac{\pi(s)}{s}e^{irs}\,ds.
\end{equation} 
Thus the $\Delta V$ function that we derive is
\begin{align}
(\Delta V)(r)&=\Delta V_{\mbox{analytic}}(ir)\nonumber\\
&=-(2\pi)^{-2}ie^2\frac{1}{ir}\int_{s=0}^{\infty}\frac{\pi(s)}{s}e^{-rs}\,ds. \nonumber\\
&=-(2\pi)^{-2}e^2\frac{1}{r}\int_{s=0}^{\infty}\frac{\pi(s)}{s}e^{-rs}\,ds. \label{eq:Delta_V_def}
\end{align}
In the case when $\pi=\pi_s$ is the spectral vacuum polarization function we have, using Eq.~\ref{eq:polarization_function}, that
\begin{align*}
(\Delta V)(r)&=-(2\pi)^{-2}e^2\frac{2}{3\pi}e^2m^3\frac{1}{r}\int_{s=2m}^{\infty}\frac{Z(s)(3+2Z^2(s))}{s^4}e^{-rs}\,ds\\
&=-(2\pi)^{-2}e^2\frac{2}{3\pi}e^2m^3\frac{1}{r}\int_{x=1}^{\infty}\frac{(x^2-1)^{\frac{1}{2}}(2x^2+1)}{(16m^4)x^4}e^{-2mxr}(2m)\,dx\\
&=-\frac{\alpha^2}{3\pi}\frac{1}{r}\int_{x=1}^{\infty}(x^2-1)^{\frac{1}{2}}(2x^2+1)x^{-4}e^{-2mxr}\,dx,
\end{align*}
where $\alpha=(4\pi)^{-1}e^2$ is the fine structure constant. (If we had chosen to analytically continue the second term of Eq.~\ref{eq:Delta_V_1} into the lower half plane and evaluate the result at $-ir$ we obtain the same result.)

The result that we have obtained is precisely the Uehling potential function which has previously only been calculated through negotiating infinities and divergences using complex calculations such as the use of charge and mass renormalization. 

From this potential function the Uehling contribution to the Lamb shift can be exactly calculated according to
\begin{equation}
\Delta E=<\psi|\Delta V|\psi>=4\pi\int|\psi|^2(r)(\Delta V)(r)r^2\,dr,
\end{equation}
where $\psi$ is the H atom 2s wave function, and theory agrees with experiment to a very high order of precision.  
 
\section{The 1 loop QED running coupling \label{section:running_coupling}}

From Eqns.~\ref{eq:total_energy} and ~\ref{eq:Delta_V_def} we have that the total equivalent potential for an electron-proton system in the Born approximation is 
\begin{eqnarray}
V(r)  & = & -\frac{e^2}{4\pi r}-(2\pi)^{-2}e^2\frac{1}{r}\int_{s=0}^{\infty}\frac{\pi(s)}{s}e^{-sr}\,ds. \nonumber
\end{eqnarray}

At range $r$ the potential is equivalent to that produced by an effective charge or running coupling constant $e_r$ given by
\begin{eqnarray}
-\frac{e_r^2}{4\pi r} & = & -\frac{e^2}{4\pi r}(1+\frac{1}{\pi}\int_{s=0}^{\infty}\frac{\pi(s)}{s}e^{-sr}\,ds) \nonumber \\
    & = & -\frac{e^2}{4\pi r}(1+\frac{1}{\pi}\int_{s=0}^{\infty}\pi(\frac{s}{r})\frac{e^{-s}}{s}\,ds). \nonumber 
\end{eqnarray}
Therefore the running fine structure ``constant" at energy $\mu$ is given by
\begin{equation} \label{eq:running_coupling}
\alpha(\mu)=\alpha(0)(1+\frac{1}{\pi}\int_{s=0}^{\infty}\pi(\mu s)\frac{e^{-s}}{s}\,ds).
\end{equation}

$\alpha(0)=(4\pi)^{-1}e^2\approx1/137$ and $\alpha$ increases with increasing energy having been measured to have a value of $\alpha(\mu)\approx1/127$ for $\mu=90$ GeV. Given the explicit expression \ref{eq:running_coupling} for the running coupling it is not necessary to use the techniques of the renormalization group equation involving a beta function to investigate its behavior. 

\subsection{Determination of the behavior of the running coupling when using the renormalized vacuum polarization function $\pi=\pi_r$ \label{sub_section:running_cc_renormalization}}

In this case we have
\begin{equation}
\pi(\mu s)=\pi_r(\mu s),
\end{equation}
where
\begin{equation} \label{eq:pi_ren_spacelike}
\pi_r(s)=-\frac{\alpha}{3\pi}(\frac{1}{3}+(3-W^2)(W\mbox{arcoth}(W)-1)),
\end{equation}
and $W$ is given by
\begin{equation} 
W=W(s)=(1+\frac{4m^2}{s^2})^{\frac{1}{2}}, s\in(0,\infty),
\end{equation}
(see Appendix 1). Note that we are using $\pi_r$ as defined in the imaginary mass domain because we are considering $s$ corresponding to spacelike $q=(0,{\vct q})$. It is shown in Appendix 1 that the function defined by Eq.~\ref{eq:pi_ren_spacelike} coincides with the function $\pi$ defined by Eq. 11.2.22 of (Weinberg, 2005).  

\begin{theorem} \label{theorem:pi_r_divergent_1}
The integral Eq.~\ref{eq:Delta_V} defining $\Delta V$ is divergent for all non-zero energies when $\pi=\pi_r$.
\end{theorem}
{\bf Proof}
Let $\mu>0$ Now
\begin{equation}
W(\mu s)=(1+\frac{4m^2}{\mu^2 s^2})^{\frac{1}{2}}, \forall s>0.
\end{equation}
Therefore
\begin{equation}
s=\frac{2m}{\mu(W^2(\mu s)-1)^{\frac{1}{2}}}.
\end{equation}
Now as $s\rightarrow\infty$, $W(\mu s)\rightarrow1^{+}$. We will now show that 
\begin{equation} \label{eq:frac_limit1}
\frac{\pi(\mu s)}{s}\rightarrow\infty,
\end{equation}
as $s\rightarrow\infty$. Terms in $\pi$ that have a finite limit as $s\rightarrow\infty$ vanish in the limit of Eq.~\ref{eq:frac_limit1}. Therefore we are interested in the limiting behavior of
\[ s\mapsto\mbox{atanh}(\frac{1}{W(\mu s)})(W^2(\mu s)-1)^{\frac{1}{2}}, \] 
as $s\rightarrow\infty$. This is the same as the limiting behavior of
\[ W\mapsto\mbox{atanh}(\frac{1}{W})(W-1)^{\frac{1}{2}}(W+1)^{\frac{1}{2}}, \]
as $W\rightarrow1^{+}$, which is the same as the limiting behavior of
\[ x\mapsto \mbox{atanh}(\frac{1}{x+1})x^{\frac{1}{2}}=\frac{x^{\frac{1}{2}}}{f(x)}, \]
as $x\rightarrow0^{+}$, where 
\begin{equation}
f(x)=\frac{1}{\mbox{atanh}(\frac{1}{x+1})}.
\end{equation}
Now, $x^{\frac{1}{2}}\rightarrow0^{+},f(x)\rightarrow0^{+}$ as $x\rightarrow0^{+}$. Therefore, by L'H\^{o}pital's rule
\begin{equation}
\lim_{x\rightarrow0^{+}}\frac{x^{\frac{1}{2}}}{f(x)}=\lim_{x\rightarrow0^{+}}\frac{\frac{1}{2}x^{-\frac{1}{2}}}{f^{\prime}(x)},
\end{equation}
if the limit exists. Now
\begin{eqnarray}
\lim_{x\rightarrow0^{+}}f^{\prime}(x) & = & \lim_{x\rightarrow0^{+}}[-(\mbox{atanh}(\frac{1}{x+1}))^{-2}\frac{1}{1-(\frac{1}{x+1})^2}(-(x+1)^{-2})] \nonumber \\
    & = & \lim_{x\rightarrow0^{+}}[(\mbox{atanh}(\frac{1}{x+1}))^{-2}\frac{1}{x(x+2)}] \nonumber \\
    & = & \lim_{x\rightarrow0^{+}}\frac{g(x)}{x(x+2)}, \nonumber \nonumber \\
\end{eqnarray}
where
\begin{equation}
g(x)=(\mbox{atanh}(\frac{1}{x+1}))^{-2}.
\end{equation}
Now
\[ g^{\prime}(x)=-2(\mbox{atanh}(\frac{1}{x+1}))^{-3}(-(x+1)^{-2}). \]
Therefore
\begin{equation}
\lim_{x\rightarrow0^{+}}f^{\prime}(x)=\lim_{x\rightarrow0^{+}}\frac{2\mbox{atanh}(\frac{1}{x+1}))^{-3}}{2x+2}=0.
\end{equation}
 Thus
\begin{equation}
\lim_{x\rightarrow0^{+}}\frac{x^{\frac{1}{2}}}{f(x)}=\infty.
\end{equation}
Therefore the integrand of the integral Eq.~\ref{eq:Delta_V} is oscillatory with ever increasing amplitude and hence the integral divergent for all non zero energies.
$\Box$

Therefore it does not seem valid to try to compute $\Delta V$ in the spacelike domain by analytic continuation of $\Delta V$ in the timelike domain since the integral defining $\Delta V$ in the timelike domain is non-convergent.

This is to be compared with the work of Landau and others relating to the Landau pole or ``ghost" pole in the solution of the renormalization group equations in QED ``the possible existence of which leads to a serious contradiction with a number of general principles of the theory" (Bogoliubov and Shirkov, 1980). 

\subsection{Determination of the behavior of the running coupling when using the spectral vacuum polarization function $\pi=\pi_s$}

In this case we have
\begin{equation}
\pi(\mu s)=\pi_s(\mu s),
\end{equation}
where
\begin{equation}
\pi_s(s)=\left\{\begin{array}{l}
\frac{2}{3\pi}e^2m^3s^{-3}Z(s)(3+2Z^2(s))\mbox{ for }s>2m \\
0\mbox{ otherwise.}
\end{array}\right.
\end{equation}

As $s\rightarrow\infty$, $s^{-1}Z(s)\rightarrow(2m)^{-1}$ and so $\pi_s(s)\rightarrow(6\pi)^{-1}e^2$. Thus
\begin{equation}
\pi(\mu s)\rightarrow\frac{e^2}{6\pi}\mbox{ as }s\rightarrow{\infty},
\end{equation}
which is a finite limit. 
\begin{theorem}
The integral given by Eq.~\ref{eq:Delta_V} defining $\Delta V$ is convergent to a finite limit for all non-zero energies $\mu>0$ when $\pi=\pi_s$.
\end{theorem} \label{theorem:convergence_1}
{\bf Proof}
Consider the case when $\mu=1$. All other values of $\mu$ can be dealt with similarly. We want to show that the integral
\[ \int\mbox{sinc}(s)\pi_s(s)\,ds, \]
is convergent. It is sufficient to show that the integral
\begin{equation} \label{eq:rcc_theorem2_integral}
\int_{s=2m}^{\infty}\mbox{sinc}(s)\frac{Z^3(s)}{s^3}\, ds,
\end{equation}
is convergent. Let
\begin{equation}
L=\lim_{s\rightarrow\infty}\frac{Z^3(s)}{s^3}=\frac{1}{(2m)^3}.
\end{equation}
Then the integral \ref{eq:rcc_theorem2_integral} will converge if $\frac{Z^3(s)}{s^3}\rightarrow L$ fast enough. Let
\begin{equation}
\epsilon_n=\mbox{sup}\left\{\left|\frac{Z^3(s)}{s^3}-L\right|:s\geq2\pi (n-1)\right\} \mbox{ for }n=1,2,\ldots.
\end{equation}
If we define
\begin{equation}
I_n=\sum_{i=1}^n(I_i^{+}-I_i^{-}),
\end{equation}
where
\begin{equation}
I_i^{+}=\int_{2\pi(i-1)}^{2\pi(i-1)+\pi}\frac{\sin(s)}{s}\,ds,
\end{equation}
and
\begin{equation}
I_i^{-}=-\int_{2\pi(i-1)+\pi}^{2\pi i}\frac{\sin(s)}{s}\,ds,
\end{equation}
then, as is well known,
\begin{equation}
I_n\rightarrow\frac{\pi}{2}\mbox{ as }n\rightarrow\infty.
\end{equation}
Now define
\begin{equation}
J_i^{+}=\int_{2\pi(i-1)}^{2\pi(i-1)+\pi}\frac{\sin(s)}{s}\frac{Z^3(s)}{s^3}\,ds,
\end{equation}
\begin{equation}
J_i^{-}=-\int_{2\pi(i-1)+\pi}^{2\pi i}\frac{\sin(s)}{s}\frac{Z^3(s)}{s^3}\,ds,
\end{equation}
and let
\begin{equation}
S_n=\sum_{i=1}^n(J_i^{+}-J_i^{-}), n=1,2,\ldots.
\end{equation}
We want to show that $S_n$ converges to a finite limit as $n\rightarrow\infty$. We have
\begin{eqnarray}
S_n & \in & (\sum_{i=1}^n((L-\epsilon_i)I_i^{+}-(L+\epsilon_i)I_i^{-}),\sum_{i=1}^n((L+\epsilon_i)I_i^{+}-(L-\epsilon_i)I_i^{-})) \nonumber \\
    & = & (\sum_{i=1}^n L(I_i^{+}-I_i^{-})-\epsilon_i(I_i^{+}+I_i^{-}),\sum_{i=1}^n L(I_i^{+}-I_i^{-})+\epsilon_i(I_i^{+}+I_i^{-})) \nonumber
\end{eqnarray}
Clearly if $\epsilon_i\rightarrow0$ fast enough then $S_n$ is convergent. Well we have
\begin{eqnarray}
\left|\frac{Z^3(s)}{s^3}-L\right| & = & \left|(\frac{1}{4m^2}-\frac{1}{s^2})^{\frac{3}{2}}-\frac{1}{(2m)^3}\right| \nonumber \\
    & = & \left|\frac{(\frac{1}{4m^2}-\frac{1}{s^2})^{3}-\frac{1}{(2m)^6}}{(\frac{1}{4m^2}-\frac{1}{s^2})^{\frac{3}{2}}+\frac{1}{(2m)^3}}\right| \nonumber \\
    & \leq & (2m)^3\left|(\frac{1}{4m^2}-\frac{1}{s^2})^{3}-\frac{1}{(2m)^6}\right| \nonumber \\
    & = & \frac{(2m)^3}{s^2}\left|\frac{3}{4m^2s^2}-\frac{3}{(2m)^4}-\frac{1}{s^4}\right|, \nonumber 
\end{eqnarray}
for $s>2m$.
Now if $s$ is sufficiently large then
\begin{eqnarray}
|\frac{3}{4m^2s^2}-\frac{3}{(2m)^4}-\frac{1}{s^4}| & = & \frac{3}{(2m)^4}-\frac{3}{4m^2s}+\frac{1}{s^4} \nonumber \\
    & < & \frac{3}{(2m)^4}+\frac{1}{s^4} \nonumber \\
    & \leq & \frac{3}{(2m)^4}+\frac{1}{(2m)^4} \nonumber \\
    & = & \frac{1}{4m^4}. \nonumber
\end{eqnarray} 
 Therefore
\begin{equation}
\mbox{sup}\left\{\left|\frac{Z^3(s)}{s^3}-L\right|:s\geq a\right\}\leq\frac{2}{ma^2} \mbox{ for $a$ sufficiently large}.
\end{equation}
Hence
\begin{equation}
\epsilon_i\leq\frac{2}{m(2\pi(i-1))^2} \mbox{ for $i$ sufficiently large}.
\end{equation}
Thus
\begin{equation}
\epsilon_i(I_i^{+}+I_i^{-})\leq\frac{2}{m(2\pi(i-1))^2}\int_{2\pi(i-1)}^{2\pi i}\frac{1}{s}\,ds\leq\frac{2}{m(2\pi(i-1))^2}\frac{1}{2\pi(i-1)}2\pi,
\end{equation}
for $i$ sufficiently large. Therefore the sequence $S_n$ is convergent as $n\rightarrow\infty$.
$\Box$

Therefore, since the integral~\ref{eq:Delta_V} defining $\Delta V$ in the timelike domain is convergent for $r\in(0,\infty)$, it is valid to carry out the analytic continuation operation we have defined leading to the form of Eq.~\ref{eq:Delta_V_def} for $\Delta V$ in the spacelike domain. 

The spectral running coupling function $\mu\mapsto\alpha(\mu)$ defined by Eq.~\ref{eq:running_coupling}, with $\pi=\pi_s$, is well defined and finite for all energies $\mu$. It is analytic over $(0,\infty)$. The spectral running coupling does not exhibit a Landau pole.

It can be seen from Eq.~\ref{eq:running_coupling} that the running coupling depends on the units used for electric charge. It is natural to use Gaussian natural units in which $\hbar=c=1$ and the potential between two charges of magnitudes $e_1$ and $e_2$ is given by
\[ V(r)=\frac{e_1e_2}{r}. \]
We find that when these units are used the spectral running coupling defined by Eq.~\ref{eq:running_coupling} with $\pi=\pi_s$ agrees with experiment.

\begin{figure} 
\centering
\includegraphics[width=15cm]{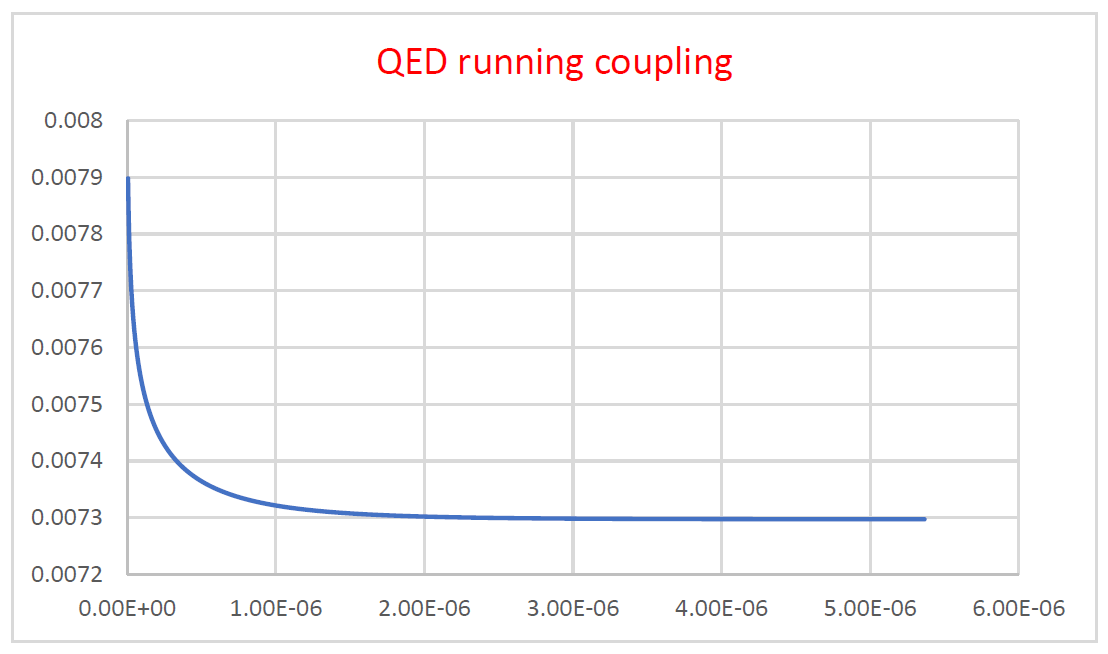}
\caption{Spectral QED running fine structure constant on the basis of vacuum polarization.} \label{fig:Fig3}
\end{figure}

A graph of the spectral running fine structure constant versus $r=\mu^{-1}$ for $r$ between 0 and $2\%$ of $a_0$ where $a_0=$ the first Bohr radius of the H atom in natural units is shown in Figure \ref{fig:Fig3}. It is seen that the running fine structure constant achieves at high energy a value of $0.0079\approx1/127$ which, up to higher order electroweak and hadronic contributions, agrees with experiment.

\section{Conclusion\label{section:conclusion}}

We have presented a spectral calculus for the computation of the spectrum of causal Lorentz invariant complex measures on Minkowski space and shown how this enables one to compute the density for such a measure with respect to Lebesgue measure. This has been applied to the case of the contraction of the vacuum polarization tensor resulting in a spectral vacuum polarization function which has very close agreement with the vacuum polarization function computed using dimensional regularization / renormalization in the domain of real mass.

Using the Born approximation together with the spectral vacuum polarization function the exact Uehling potential function is derived from which the Uehling contribution to the Lamb shift can be exactly computed.
 
With the spectral vacuum polarization function we obtain a well defined convergent running coupling function whereas the integral defining the running coupling function generated using dimensional regularization / renormalization is shown to be divergent at all non-zero energies. 

In subsequent work we will apply the spectral calculus to the electron self energy and generally to all renormalization issues arising in the QFT of the electroweak force. In addition QCD will be formulated in the context of locally conformally flat space-times (M\"{o}bius structures) (Mashford, 2017a and b) and the running coupling constant for QCD will be computed with a view to proving, or deriving, the asymptotic freedom of QCD.

\section*{References}

\rf Bogoliubov, N. N., Logunov, A. A. and Todorov, I. T., {\em Introduction to Axiomatic Quantum Field Theory}, Benjamin, 1975.

\rf Bogoliubov, N. N., and Shirkov, D. V., {\em Introduction to the theory of quantized fields}, Third edition, Wiley, New York, 1980.

\rf Choquet-Bruhat, Y., DeWitte-Morette, C., and Dillard-Bleick, M., {\em Analysis, manifolds and physics}, North-Holland, Amsterdam, 1982.

\rf Colombeau, J. F., {\em New generalized functions and multiplication of distributions}, North Holland, Amsterdam, 1984.

\rf Epstein, H. and Glaser, V., ``Role of locality in perturbation theory", {\em Annales de l'Institut Henri Poincar\'{e} Section A Physique Th\'{e}orique} 19(3), 211-295, 1973.

\rf Halmos, P. R., {\em Measure theory}, Springer-Verlag, New York, 1988.

\rf Itzykson, C. and Zuber, J.-B., {\em Quantum Field Theory}, McGraw-Hill, New York, 1980.

\rf Mandl, F. and Shaw, G., {\em Quantum Field Theory}, Wiley, Chichester, 1991.

\rf Mashford, J. S., ``A non-manifold theory of space-time", {\em Journal of Mathematical Physics} 22(9), 1981, 1990-1993.

\rf Mashford, J. S., {\em Invariant  measures and M\"{o}bius structures: A framework for field theory}, PhD thesis, University of Melbourne, 2005.

\rf Mashford, J, ``An approach to classical quantum field theory based on the geometry of locally conformally flat space-time," \textit{Advances in Mathematical Physics}, vol. 2017, Article ID 8070462, 2017a.

\rf Mashford J. S., ``Second quantized quantum field theory based on invariance properties of locally conformally flat space-times",  arXiv:1709.09226, 2017b.

\rf Oberguggenberger, M., {\em Multiplication of distributions and applications to partial differential equations}, Longman, 1992.

\rf Pohl, R., ``Laser Spectroscopy of Muonic Hydrogen and the Puzzling Proton", Journal of the Physical Society of Japan 85, 091003 (2016).

\rf Scharf, G., {\em Finite Quantum Electrodynamics: The Causal Approach}, Springer-Verlag, Berlin, 1995.

\rf Weinberg, S., {\em The quantum theory of fields}, Vol. 1, Cambridge University Press, 2005.

\rf Wilson, K., ``The renormalization group and critical phenomena", {\em Reviews of Modern Physics}, 55(3), 1983.

\section*{Appendix 1: Derivation of closed form solution forregularized/renormalized vacuum polarization}

The standard formula for the vacuum polarization function $\pi_r$ as obtained using regularization and renormalization is
\begin{equation} \label{eq:appendix1_vp1}
\pi_r(k^2)=-\frac{2\alpha}{\pi}\int_0^1d\beta\,\beta(1-\beta)\log\left(1-\frac{k^2\beta(1-\beta)}{m^2}\right),
\end{equation}
(Mandl and Shaw, 1991, p. 229) where $m>0$ is the mass of the electron and (in natural units) $\alpha=(4\pi)^{-1}e^2$ is the fine structure constant in which $e>0$ is the magnitude of the charge of the electron. $\pi_r$ is defined for all $k\in{\bf R}^4$ for which $k^2<4m^2$. This integral can be performeed leading to the closed form solution (see (Itzykson and Zuber, 1980, p. 323)) 
\begin{equation} \label{eq:pi_Itzykson}
\pi_r(k^2)=-\frac{\alpha}{3\pi}\left\{\frac{1}{3}+2\left(1+\frac{2m^2}{k^2}\right)\left[\left(\frac{4m^2}{k^2}-1\right)^{\frac{1}{2}}\mbox{arcot}\left(\frac{4m^2}{k^2}-1\right)^{\frac{1}{2}}-1\right]\right\}.
\end{equation}
The function defined by Eq.~\ref{eq:pi_Itzykson} is only defined for $0<k^2<4m^2$ unless one allows the functions $z\mapsto z^{\frac{1}{2}}$ and $z\mapsto\mbox{arcot}(z)$ to be defined on complex domains. It is useful for the work of the present paper to write down the derivation of this result and to consider the answer when $k^2<0$.

Let $m>0$ and $I:\{k\in{\bf R}^4:k^2<4m\}\rightarrow(-\infty,0)$ be defined by
\begin{equation}
I(k)=2\int_{\beta=0}^1d\beta\,\beta(1-\beta)\log(1-\frac{k^2\beta(1-\beta)}{m^2}),
\end{equation}
Then
\begin{eqnarray}
I(k) & = & 2\int_{\beta=0}^1d(\frac{1}{2}\beta^2-\frac{1}{3}\beta^3)\log(1-\frac{k^2\beta(1-\beta)}{m^2}) \nonumber \\
    & = & -2\int_{\beta=0}^1\frac{\frac{1}{2}\beta^2-\frac{1}{3}\beta^3}{1-\beta(1-\beta)m^{-2}k^2}\frac{k^2}{m^2}(2\beta-1)\,d\beta \nonumber \\
    & = & 2\int_{\beta=0}^1\frac{(\frac{1}{3}\beta^3-\frac{1}{2}\beta^2)(2\beta-1)}{m^2(k^2)^{-1}-\beta(1-\beta)}\,d\beta \nonumber
\end{eqnarray}
Now
\begin{equation}
\frac{m^2}{k^2}-\beta(1-\beta)=\frac{m^2}{k^2}-\beta+\beta^2=(\beta-\frac{1}{2})^2-\frac{1}{4}+\frac{m^2}{k^2}.
\end{equation}
Therefore, changing variables,
\begin{eqnarray}
I(k) & = & 2\int_{\beta=-\frac{1}{2}}^{\frac{1}{2}}\frac{(\frac{1}{3}(\beta+\frac{1}{2})^3-\frac{1}{2}(\beta+\frac{1}{2})^2)2\beta}{\beta^2-\frac{1}{4}+m^2(k^2)^{-1}}\,d\beta \label{eq:appendix1_vp2} \\
    & = & 2\int_{\beta=-\frac{1}{2}}^{\frac{1}{2}}\frac{(\frac{1}{3}(\beta+\frac{1}{2})^3-\frac{1}{2}(\beta+\frac{1}{2})^2)2\beta}{\beta^2+X^2}\,d\beta
\end{eqnarray}
where
\begin{equation}
X=X(k)=\frac{1}{2}(\frac{4m^2}{k^2}-1)^{\frac{1}{2}}\in(0,\infty), \mbox{ for } 0<k^2<4m^2.
\end{equation}
Let $\beta=X\tan(u)$. Then $\beta^2+X^2=X^2\sec^2(u)$ and $d\beta=X\sec^2(u)\,du$. Also
\begin{equation}
\frac{1}{3}(\beta+\frac{1}{2})^3-\frac{1}{2}(\beta+\frac{1}{2})^2=\frac{1}{3}(\beta+\frac{1}{2})^2(\beta-1).
\end{equation}
Thus 
\begin{eqnarray}
I(k) & = & \frac{4}{3}\frac{1}{X}\int_{\beta=-\frac{1}{2}}^{\frac{1}{2}}(\beta+\frac{1}{2})^2(\beta-1)\beta\,du \nonumber \\
    & = & \frac{4}{3}\frac{1}{X}\int_{\beta=-\frac{1}{2}}^{\frac{1}{2}}\beta^4-\frac{3}{4}\beta^2\,du,
\end{eqnarray}
(the integral of the odd powers of $\beta$ vanishes). Therefore
\begin{eqnarray}
I(k) & = & \frac{4}{3}\frac{1}{X}\int_{\beta=-\frac{1}{2}}^{\frac{1}{2}}X^4\tan^4(u)-\frac{3}{4}X^2\tan^2(u)\,du \nonumber \\
    & = & \frac{4}{3}(X^3(\frac{1}{3}\tan^3(u)-\tan(u)+u)-\frac{3}{4}X(\tan(u)-u))\left|_{\beta=-\frac{1}{2}}^\frac{1}{2}\right. \nonumber \\
    & = & \frac{1}{3}(\frac{1}{3}+(4X^2+3)(2X\mbox{arcot}(2X)-1)), \label{eq:appendix1_vp_result1}
\end{eqnarray}
which leads directly to the required result.

Clearly, when $k^2=0$, $I(k)$ defined by Eq.~\ref{eq:appendix1_vp1} has the value $I(k)=0$. Now consider the case when $k^2<0$. Then we proceed with the same steps up to Eq.~\ref{eq:appendix1_vp2} but now we write
\begin{equation}
I(k)=2\int_{\beta=-\frac{1}{2}}^{\frac{1}{2}}\frac{(\frac{1}{3}(\beta+\frac{1}{2})^3-\frac{1}{2}(\beta+\frac{1}{2})^2)2\beta}{{\beta^2-W^2}}\,d\beta,
\end{equation}
where
\begin{equation} \label{eq:appendix1_W_def}
W=W(k)=(\frac{1}{4}+\frac{m^2}{-k^2})^{\frac{1}{2}}=\frac{1}{2}(1-\frac{4m^2}{k^2})^{\frac{1}{2}}.
\end{equation}
Then make the substitution $\beta=W\mbox{tanh}(u)$ so that
\[ \beta^2-W^2=-W^2\mbox{sech}^2(u), d\beta=W\mbox{sech}^2(u)\,du. \]
Therefore
\begin{eqnarray}
I(k) & = & -\frac{4}{3}\frac{1}{W}\int_{\beta=-\frac{1}{2}}^{\frac{1}{2}}\beta^4-\frac{3}{4}\beta^2\,du, \nonumber \\
       & = & -\frac{4}{3}\frac{1}{W}\int_{\beta=-\frac{1}{2}}^{\frac{1}{2}}W^4\mbox{tanh}^4(u)-\frac{3}{4}W^2\mbox{tanh}^2(u)\,du \nonumber
\end{eqnarray}
Now
\[ \int\mbox{tanh}^2(u)\,du=u-\mbox{tanh}(u)+c, \]
\[ \int\mbox{tanh}^4(u)\,du=u-\mbox{tanh}(u)-\frac{1}{3}\mbox{tanh}^3(u)+c. \]
Thus
\begin{eqnarray}
I(k) & = & -\frac{4}{3}\frac{1}{W}\int_{\beta=-\frac{1}{2}}^{\frac{1}{2}}W^4\mbox{tanh}^4(u)-\frac{3}{4}W^2\mbox{tanh}^2(u)\,du \nonumber \\
    & = & -\frac{4}{3}(W^3(-\frac{1}{3}\mbox{tanh}^3(u)-\mbox{tanh}(u)+u)-\frac{3}{4}W(-\mbox{tanh}(u)+u))\left|_{\beta=-\frac{1}{2}}^\frac{1}{2}\right. \nonumber \\
    & = & \frac{1}{3}(\frac{1}{3}+(3-4W^2)(2W\mbox{arcoth}(2W)-1)). \label{eq:appendix_vp_result2}
\end{eqnarray}
This result may be obtained more directly by noting that
\begin{equation}
X=\frac{1}{2}\left(\frac{4m^2}{k^2}-1\right)^{\frac{1}{2}}=i\frac{1}{2}\left(1-\frac{4m^2}{k^2}\right)^{\frac{1}{2}}=iW,
\end{equation}
when $k^2<0$ and
\begin{equation}
\mbox{arcot}(2X)=\mbox{atan}(\frac{1}{2X})=-\mbox{atan}(\frac{i}{2W})=-i\mbox{atanh}(\frac{1}{2W})=-i\mbox{arcoth}(2W).
\end{equation}
and then using Eq.~\ref{eq:appendix1_vp_result1}. Thus the renormalized vacuum polarization when $k^2<0$ is given by
\begin{equation} 
\pi_r(k^2)=-\frac{\alpha}{3\pi}(\frac{1}{3}+(3-4W^2)(2W\mbox{arcoth}(2W)-1)),
\end{equation}
where $W:\{k\in{\bf R}^4:k^2<0\}\rightarrow(0,\infty)$ is given by Eq.~\ref{eq:appendix1_W_def}.

Thus, in other words, 
\begin{equation} \label{eq:pi_Itzykson1}
\pi_r(k^2)=-\frac{\alpha}{3\pi}(\frac{1}{3}+(3-W^2)(W\mbox{arcoth}(W)-1)),
\end{equation}
where $W$ is given by
\begin{equation} \label{eq:appendix1_W_def1}
W=W(k)=(1-\frac{4m^2}{k^2})^{\frac{1}{2}}.
\end{equation}

\section*{Appendix 2: Proof of the Spectral Theorem}

The action of the proper orthochronous Lorentz group $O(1,3)^{+\uparrow}$ on Minkowski space has 5 classes of orbits each corresponding to a particular isotropy subgroup (little group). Firstly there is the distinguished orbit $\{0\}$ consisting of the origin. Secondly there are the positive mass hyperboloids $\{p\in{\bf R}^4:p^2=m^2, p^0>0\}$ with little group isomorphic to $SO(3)$. Then there are the negative mass hyperboloids, the positive open null cone, the negative open null cone and the imaginary mass hyperboloids. The spectral theorem is proved by considering separately each class of orbit. We will prove it for the space $X=\{p\in{\bf R}^4:p^2>0,p^0>0\}$ consisting of the union of all positive mass hyperboloids. The other cases can be proved similarly. We will prove the spectral theorem first for Lorentz invariant measures $\mu:{\mathcal B}_0(X)\rightarrow[0,\infty]$ and then generalize the theorem later to Lorentz invariant complex measures.

Let Rotations $\subset O(1,3)^{+\uparrow}$ be defined by
\begin{equation}
\mbox{Rotations}=\left\{\left(\begin{array}{ll}
1 & 0 \\
0 & A
\end{array}\right):A\in SO(3)\right\},
\end{equation}
and Boosts $\subset O(1,3)^{+\uparrow}$ be the set of pure boosts. Then it can be shown that for every $\Lambda\in O(1,3)^{+\uparrow}$ there exist unique $B\in$ Boosts and $R\in$ Rotations such that
\[ \Lambda = BR. \]
Thus there exist maps $\pi_1:O(1,3)^{+\uparrow}\rightarrow$ Boosts and $\pi_2:O(1,3)^{+\uparrow}\rightarrow$ Rotations such that for all $\Lambda\in O(1,3)^{+\uparrow}$
\begin{eqnarray}
&  & \Lambda = \pi_1(\Lambda)\pi_2(\Lambda), \nonumber \\
&  & \Lambda = BR \mbox{ with } B\in\mbox{Boosts and }R\in\mbox{ Rotations }\Rightarrow B=\pi_1(\Lambda), R=\pi_2(\Lambda). \nonumber
\end{eqnarray}
For $m>0$, define $h_m:\mbox{Boosts}\rightarrow H_m$ by
\begin{equation}
h_m(B)=B(m,{\vct 0})^{T}.
\end{equation}
We will show that $h_m$ is a bijection. Let $p\in H_m$. Choose $\Lambda\in O(1,3)^{+\uparrow}$ such that $p=\Lambda(m,{\vct 0})^{T}$. Then $p=\pi_1(\Lambda)\pi_2(\Lambda)(m,{\vct 0})^{T}=\pi_1(\Lambda)(m,{\vct 0})^{T}\in h(\mbox{Boosts})$. Therefore $h_m$ is surjective. Now suppose that $h(B_1)=h(B_2)$. Then $B_1(m,{\vct 0})^{T}=B_2(m,{\vct 0})^{T}$. Thus $B_2^{-1}B_1(m,{\vct 0})^{T}=(m,{\vct 0})^{T}$. Hence $B_2^{-1}B_1=R$ for some $R\in$ Rotations. Therefore $B_1=\pi_1(B_1)=\pi_1(B_2R)=\pi_1(B_2)=B_2$. Therefore $h_m$ is a bijection.

Now there is an action $\rho_m:O(1,3)^{+\uparrow}\times H_m\rightarrow H_m$ of $O(1,3)^{+\uparrow}$ on $H_m$ defined by
\begin{equation}
\rho_m(\Lambda,p)=\Lambda p.
\end{equation}
$\rho_m$ induces an action ${\tld \rho}_m:O(1,3)^{+\uparrow}\times$Boosts$\rightarrow$Boosts according to
\begin{eqnarray}
{\tld \rho}_m(\Lambda,B) & = & h_m^{-1}(\rho_m(\Lambda,h_m(B))) \nonumber \\
    & = & h_m^{-1}(\Lambda B(m,{\vct 0})^{T}) \nonumber \\
    & = & h_m^{-1}(\pi_1(\Lambda B)\pi_2(\Lambda B)(m,{\vct 0})^{T}) \nonumber \\
    & = & h_m^{-1}(\pi_1(\Lambda B)(m,{\vct 0})^{T}) \nonumber \\
    & = & \pi_1(\Lambda B).
\end{eqnarray}
Note that the induced action is independent of $m$ for all $m>0$.

 Let
\begin{equation}
X=\bigcup_{m>0}H_m=\{p\in{\bf R}^4:p^2>0,p^0>0\}.
\end{equation}
Define the action $\rho:O(1,3)^{+\uparrow}\times X\rightarrow X$ by $\rho(\Lambda,p)=\Lambda p$. Then $\rho$ induces an action ${\tld\rho}:O(1,3)^{+\uparrow}\times(0,\infty)\times\mbox{Boosts}\rightarrow(0,\infty)\times\mbox{Boosts}$ according to
\begin{equation} \label{eq:action_Boosts}
{\tld\rho}(\Lambda,m,B)={\tld\rho}_m(B)=\pi_1(\Lambda B).
\end{equation}
Defime, for each $m>0$, $f_m:O(1,3)^{+\uparrow}\rightarrow H_m\times\mbox{Rotations}$ by
\begin{equation}
f_m(\Lambda)=(h_m(\pi_1(\Lambda)),\pi_2(\Lambda)).
\end{equation}
Then each $f_m$ is a bijection.
The map $g:(0,\infty)\times\mbox{Boosts}\rightarrow X$ defined by
\begin{equation}
g(m,B)=h_m(B)=B(m,{\vct 0})^{T},
\end{equation}
is a bijection. Define $f:(0,\infty)\times O(1,3)^{+\uparrow}\rightarrow X\times\mbox{Rotations}$ by
\begin{equation}
f(m,\Lambda)=f_m(\Lambda).
\end{equation}
$f$ is a bijection so we can push forward or pull back measures using $f$ at will. 

Suppose that $\mu:{\mathcal B}_0(X)\rightarrow[0,\infty]$ is a measure on $X$ and that $\mu$ is invariant under the action. Let $\mu_R$ be the measure on Rotations induced by Haar measure on $SO(3)$. Let $\nu$ be the product measure $\nu=\mu\times\mu_R$ whose existence and uniqueness is guaranteed by the Hahn-Kolmogorov theorem and the fact that both $X$ and Rotations are $\sigma$-finite. Let $\nu\#f^{-1}$ denote the pull back of $\nu$ by $f$ (i.e. the push forward of $\nu$ by $f^{-1}$). Then
\begin{equation}
(\nu\#f^{-1})(\Gamma)=\nu(f(\Gamma)), \forall\Gamma\in{\mathcal B}_0((0,\infty)\times O(1,3)^{+\uparrow}).
\end{equation}
Consider the action $\tau:O(1,3)^{+\uparrow}\times(0,\infty)\times O(1,3)^{+\uparrow}\rightarrow(0,\infty)\times O(1,3)^{+\uparrow}$ defined by
\begin{equation}
\tau(\Lambda,m^{\prime},\Lambda^{\prime})=(m^{\prime},\Lambda\Lambda^{\prime}).
\end{equation}
$\tau$ induces an action ${\tld \tau}:O(1,3)^{+\uparrow}\times X\times\mbox{Rotations}\rightarrow X\times\mbox{Rotations}$ so that if $p^{\prime}\in X$ with $p^{\prime}=B^{\prime}(m^{\prime},{\vct 0})^{T}$ with $m^{\prime}\in(0,\infty), B^{\prime}\in\mbox{Boosts}$ and $R^{\prime}\in\mbox{Rotations}$ then
\begin{eqnarray}
{\tld \tau}(\Lambda,(p^{\prime},R^{\prime})) & = & {\tld \tau}(\Lambda,B^{\prime}(m^{\prime},{\vct 0})^{T},R^{\prime}) \nonumber \\
    & = & f_{m^{\prime}}(\Lambda f_{m^{\prime}}^{-1}(h_{m^{\prime}}(B^{\prime}(m^{\prime},{\vct 0})^{T},R^{\prime})) \nonumber \\
    & = & f_{m^{\prime}}(\Lambda f_{m^{\prime}}^{-1}(h_{m^{\prime}}(\pi_1(\Lambda^{\prime})),\pi_2(\Lambda^{\prime})) \nonumber \\
    & = & f_{m_{\prime}}(\Lambda\Lambda^{\prime}) \nonumber \\
    & = & (h_{m^{\prime}}(\pi_1(\Lambda\Lambda^{\prime})),\pi_2(\Lambda\Lambda^{\prime})), \nonumber
\end{eqnarray}
where $\Lambda^{\prime}=B^{\prime}R^{\prime}$. We will now show that the measure $\nu$ is an invariant measure on $X\times\mbox{Rotations}$ with respect to the action ${\tld \tau}$. To this effect let $E_1^{\prime}\subset\mbox{Boosts}, E_2^{\prime}\subset\{(m^{\prime},{\vct 0}):m^{\prime}\in(0,\infty)\}$ and $E_3^{\prime}\subset\mbox{Rotations}$ be Borel sets. Then
\begin{eqnarray}
\nu({\tld\tau}(\Lambda,E_1^{\prime}E_2^{\prime},E_3^{\prime})) & = & \nu(\pi_1(\Lambda E_1^{\prime})E_2^{\prime}\times\pi_2(\Lambda E_3^{\prime})) \nonumber \\
    & = & \mu(\pi_1(\Lambda E_1^{\prime})E_2^{\prime})\mu_R(\pi_2(\Lambda E_3^{\prime})) \nonumber \\
    & = & \mu(\pi_1(\Lambda E_1^{\prime}))\mu_R(\pi_2(\Lambda)\pi_2(E_3^{\prime})) \nonumber \\
    & = & \mu(E_1^{\prime}E_2^{\prime})\mu_R(E_3^{\prime}) \nonumber \\
    & = & \nu(E_1^{\prime}E_2^{\prime},E_3^{\prime}), \nonumber
\end{eqnarray}
 (here we have used the notation of juxtaposition of sets to denote the set of all products i.e. $ S_1S_2=\{xy:x\in S_1, y\in S_2\}$, also $xS=\{xy:y\in S\}$). Therefore the measure $\nu\#f^{-1}$ is an invariant measure on $O(1,3)^{+\uparrow}$ with respect to the action $\tau$. Therefore for each Borel set $E\subset(0,\infty)$ the measure $(\nu\#f^{-1})_E:{\mathcal B}_0(O(1,3)^{+\uparrow})\rightarrow[0,\infty]$ defined by
\begin{equation}
(\nu\#f^{-1})_E(\Gamma)=(\nu\#f^{-1})(E,\Gamma),
\end{equation}
is a translation invariant measure on the group $O(1,3)^{+\uparrow}$. Therefore since, $O(1,3)^{+\uparrow}$ is a locally compact second countable topological group there exists, by the uniqueness part of Haar's theorem, a unique $c=c(E)\ge0$ such that 
\begin{equation}
(\nu\#f^{-1})_E=c(E)\mu_{O(1,3)^{+\uparrow}},
\end{equation} 
where $\mu_{{O(1,3}^{+\uparrow}}$ is the Haar measure on $O(1,3)^{+\uparrow}.$ Denote by $\sigma$ the map $\sigma:{\mathcal B}_0((0,\infty))\rightarrow[0,\infty]$ defined by $\sigma(E)=c(E)$.

We will now show that $\sigma$ is a measure on $(0,\infty)$. We have that for any $\Gamma\in{\mathcal B}_0(O(1,3)^{+\uparrow}),E\in{\mathcal B}_0((0,\infty))$
\begin{equation}
\sigma(E)\mu_{O(1,3)^{+\uparrow}}(\Gamma)=(\nu\#f^{-1})_E(\Gamma)=\nu(f(E,\Gamma))=\nu(\pi_1(\Gamma)E\times\pi_2(\Gamma)).
\end{equation}
Choose $\Gamma\in{\mathcal B}_0(O(1,3)^{+\uparrow})$ such that $\mu_{O(1,3)^{+\uparrow}}(\Gamma)\in(0,\infty)$.

Then 
\begin{equation}
\sigma(\emptyset)\mu_{O(1,3)^{+\uparrow}}(\Gamma)=\nu(\pi_1(\Gamma)\emptyset\times\pi_2(\Gamma))=\nu(\emptyset)=0.
\end{equation}
Therefore 
\begin{equation}
\sigma(\emptyset)=0.
\end{equation}
Also let $\{E_n\}_{n=1}^{\infty}\subset{\mathcal B}_0((0,\infty))$.  Then
\begin{eqnarray}
\sigma(\bigcup_{n=1}^{\infty}(E_n) & = & \mu_{O(1,3)^{+\uparrow}}(\Gamma)^{-1}\nu(\pi_1(\Gamma)\bigcup_{n=1}^{\infty}E_n\times\pi_2(\Gamma)) \nonumber \\
     & = & \mu_{O(1,3)^{+\uparrow}}(\Gamma)^{-1}\nu(\bigcup_{n=1}^{\infty}(\pi_1(\Gamma)E_n\times\pi_2(\Gamma))) \nonumber \\
     & = & \sum_{n=1}^{\infty}\mu_{O(1,3)^{+\uparrow}}(\Gamma)^{-1}\nu((\pi_1(\Gamma)E_n\times\pi_2(\Gamma))) \nonumber \\
     & = & \sum_{n=1}^{\infty}\sigma(E_n).
\end{eqnarray}
Thus $\sigma$ is a measure. 

The above argument holds for all invariaant measures $\mu:{\mathcal B}_0(X)\rightarrow[0,\infty]$. Therefore, in particular, it is true for $\Omega_m$ for $m\in(0,\infty)$. Hence there exists a measure $\sigma_{\Omega_m}:{\mathcal B}_0((0,\infty))\rightarrow[0,\infty]$ such that
\begin{equation}
((\Omega_m\times\mu_{SO(3)})\#f^{-1})(E,\Gamma)=\sigma_{\Omega_m}(E)\mu_{O(1,3)^{+\uparrow}}(\Gamma),
\end{equation}
for $E\in{\mathcal B}_0((0,\infty)), \Gamma\in{\mathcal B}_0(O(1,3)^{+\uparrow}$. But
\begin{eqnarray}
((\Omega_m\times\mu_{SO(3)})\#f^{-1})(E,\Gamma) & = & (\Omega_m\times\mu_{SO(3)})(f(E,\Gamma)) \nonumber \\
    & = &  (\Omega_m\times\mu_{SO(3)})(\pi_1(\Gamma)(E\times\{{\vct 0}\}),\pi_2(\Gamma)) \nonumber \\
    & = & \Omega_m(\pi_1(\Gamma)(E\times\{{\vct 0}\}))\mu_{SO(3)}(\pi_2(\Gamma)) \nonumber \\
    & = & \Omega_m(\pi_1(\Gamma)(m,{\vct 0})^{T})\mu_{SO(3)}(\pi_2(\Gamma))\delta_E(m) \nonumber \\
\end{eqnarray}
where $\delta_m$ is the Dirac measure concentrated on $m$. Thus
\begin{equation}
\sigma_{\Omega_m}(E)=\mu_{O(1,3)^{+\uparrow}}(\Gamma)^{-1}\Omega_m(\pi_1(\Gamma)(m,{\vct 0})^{T})\mu_{SO(3)}(\pi_2(\Gamma))\delta_m(E),
\end{equation}
for any $E\in{\mathcal B}_0((0,\infty)),\Gamma\in{\mathcal B}_0(O(1,3)^{+\uparrow})$ such that $\mu_{O(1,3)^{+\uparrow}}(\Gamma)\in(0,\infty).$ Choose any $\Gamma\in{\mathcal B}_0(O(1,3)^{+\uparrow})$ such that $\mu_{O(1,3)^{+\uparrow}}(\Gamma)\in(0,\infty)$ and define $\sigma_{\Omega}:(0,\infty)\rightarrow(0,\infty)$ by
\begin{equation}
\sigma_{\Omega}(m)=\mu_{O(1,3)^{+\uparrow}}(\Gamma)^{-1}\Omega_m(\pi_1(\Gamma)(m,{\vct 0})^{T})\mu_{SO(3)}(\pi_2(\Gamma)).
\end{equation}
Then
\begin{equation}
\sigma_{\Omega_m}=\sigma_{\Omega}(m)\delta_m, \forall m\in(0,\infty).
\end{equation}

Returning now to the general invariant measure $\mu:{\mathcal B}_0(X)\rightarrow[0,\infty]$ we will now show that $\mu$ can be written as a product $\mu=\sigma\times\mu_B$ for some measure $\mu_B:{\mathcal B}_0(\mbox{Boosts})\rightarrow[0,\infty]$ relative to the identification $g:(0,\infty)\times\mbox{Boosts}\rightarrow X$. We have
\begin{equation}
\nu(\Gamma\times F)=\mu(\Gamma)\mu_{SO(3)}(F), \forall\Gamma\in{\mathcal B}_0(X),F\in{\mathcal B}(SO(3)).
\end{equation}
Therefore
\begin{equation}
\mu(\Gamma)=\mu_{SO(3)}(F)^{-1}\nu(\Gamma\times F),\forall\Gamma\in{\mathcal B}_0(X),F\in{\mathcal B}(SO(3)),\mbox{ such that }\mu_{SO(3)}(F)>0.
\end{equation}
Choose $F\in{\mathcal B}(SO(3)),\mbox{ such that }\mu_{SO(3)}(F)>0$. Then for all $\Gamma\in{\mathcal B}_0(X)$
\begin{eqnarray}
\mu(\Gamma) & = & \mu_{SO(3)}(F)^{-1}\nu(\Gamma\times F) \nonumber \\
    & = & \nu_{SO(3)}(F)^{-1}\nu(f(f^{-1}(\Gamma\times F))) \nonumber \\
    & = & \mu_{SO(3)}(F)^{-1}\nu(f(f^{-1}(B(E\times\{{\vct 0}\})\times F))) \nonumber \\
    & = & \mu_{SO(3)}(F)^{-1}\nu(f((E\times\{{\vct 0}\})\times BF)) \nonumber \\
    & = & \mu_{SO(3)}(F)^{-1}\sigma(E)\mu_{O(1,3)^{+\uparrow}}(BF) \nonumber \\
    & = & \sigma(E)\mu_B(B), \nonumber
\end{eqnarray} 
where $\Gamma=g(E\times B)=B(E\times\{{\vct 0}\})$ and
\begin{equation}
\mu_B(B)=\mu_{SO(3)}(F)^{-1}\mu_{O(1,3)^{+\uparrow}}(BF).
\end{equation}
It is straightforward to show that $\mu_B$ is a well defined measure.

Therefore for any measurable function $\psi:X\rightarrow[0,\infty]$
\begin{eqnarray}
<\mu,\psi> & = & \int\psi(p)\,\mu(dp)\nonumber \\
    & = & \int\psi(g(m,B))\,\mu_B(dB)\,\sigma(dm) \nonumber \\
    & = & \int<M_m,\psi>\,\sigma(dm), \label{eq:spectral_rep}
\end{eqnarray}
where
\begin{equation}
<M_m,\psi>=\int\psi(g(m,B))\mu_B(dB).
\end{equation}
It is straightforward to show that for all $m\in(0,\infty)$ $M_m$ defines a measure on $X$ with supp$(M_m)=H_m$. Therefore by the above argument, there exists $c=c(m)\in(0,\infty)$ such that $M_m=c_m\Omega_m$. This fact, together with the spectral representation Eq.~\ref{eq:spectral_rep} establishes (rescaling $\sigma$) that there exists a measure $\sigma:{\mathcal B}_0((0,\infty))\rightarrow[0,\infty]$ such that
\begin{equation} \label{eq:spectral_decomposition1}
\mu(\Gamma)=\int_{m=0}^{\infty}\Omega_m(\Gamma)\,\sigma(dm),
\end{equation}
as desired. 

Now suppose that $\mu:{\mathcal B}_0(X)\rightarrow{\bf R}$ is a signed measure which is Lorentz invariant. Then by the Jordan decomposition theorem $\mu$ has a decomposition $\mu=\mu^{+}-\mu^{-}$ where $\mu^{+},\mu^{-}:{\mathcal B}_0({\bf R}^4)\rightarrow[0,\infty]$ are measures. $\mu^{+}$ and $\mu^{-}$ must be Borel (finite on compact sets). In fact if $P,N\in{\mathcal B}_0(X)$ is a Hahn decomposition of $X$ with respect to $\mu$ then
\begin{equation}
\mu^{+}(\Gamma)=\mu(\Gamma\cap P), \mu^{-}(\Gamma)=\mu(\Gamma\cap N), \forall\Gamma\in{\mathcal B}_0(X).
\end{equation}
Now let $\Lambda\in O(1,3)^{+\uparrow}$ Then since $(\Lambda P)\cup(\Lambda N)=\Lambda(P\cup N)=X$, $(\Lambda P)\cap (\Lambda N)=\Lambda(P\cap N)=\emptyset$, $\mu((\Lambda P)\cap\Gamma)=\mu((\Lambda P)\cap(\Lambda\Lambda^{-1}\Gamma))=\mu(P\cap(\Lambda^{-1}\Gamma))\ge0$ and, similarly, $\mu((\Lambda N)\cap\Gamma)\le0$ $\Lambda P$ and $\Lambda N$ form a Hahn decomposition of $\mu$. Therefore 
\begin{equation}
\mu^{+}(\Lambda\Gamma)=\mu(\Lambda P\cap(\Lambda\Gamma))=\mu(\Lambda(P\cap\Gamma))=\mu(P\cap\Gamma)=\mu^{+}(\Gamma).
\end{equation}
Hence $\mu^{+}$ is a Lorentz invariant measure. Therefore it has a spectral decomposition of the form of Eq.~\ref{eq:spectral_decomposition1}. Sinilarly $\mu^{-}$ is a Lorentz invariant measure and so it has a spectral decomposition of the form of Eq.~\ref{eq:spectral_decomposition1}. Thus $\mu$ has a spectral decomposition of the form of Eq.~\ref{eq:spectral_decomposition1} where $\sigma:{\mathcal B}_0((0,\infty)\rightarrow{\bf R}$ is a signed measure. 

Finally suppose that $\mu:{\mathcal B}_0(X)\rightarrow{\bf C}$ is a Lorentz invariant complex measure. Define Re$(\mu):{\mathcal B}_0(X)\rightarrow{\bf R}$ and Im$(\mu):{\mathcal B}_0(X)\rightarrow{\bf R}$ by
\begin{equation}
(\mbox{Re}(\mu))(\Gamma)=\mbox{Re}(\mu(\Gamma)), (\mbox{Im}(\mu))(\Gamma)=\mbox{Im}(\mu(\Gamma)), \forall\Gamma\in{\mathcal B}_0(X).
\end{equation}
Then for all $\Lambda\in O(1,3)^{+\uparrow}$
\begin{equation}
(\mbox{Re}(\mu))(\Lambda\Gamma)=\mbox{Re}(\mu(\Lambda\Gamma))=\mbox{Re}(\mu(\Gamma))=(\mbox{Re}(\mu))(\Gamma).
\end{equation}
Thus Re$(\mu)$ is a Lorentz invariant signed measure and so has a representation of the form of Eq.~\ref{eq:spectral_decomposition1} for some signed measure $\sigma$. Similarly Im$(\mu)$ has such a representation. Therefore $\mu$ has a representation of this form for some complex spectral measure $\sigma:{\mathcal B}_0((0,\infty))\rightarrow{\bf C}$. This completes the proof of the spectral theorem.

\section*{Appendix 3: Dirac spinors}

\subsection*{Construction of the Dirac spinors}

Dirac spinors are usually obtained by seeking solutions to the Dirac equation of the form
\begin{eqnarray}
\psi^{+}(x) & = & e^{-ik.x}u(k), \mbox{ positive energy} \nonumber \\
\psi^{-}(x) & = & e^{ik.x}v(k), \,\,\,\mbox{ negative energy},
\end{eqnarray}
(Itzykson and Zuber, 1980, p. 55). 

Thus, in general, we are seeking solutions to the Dirac equation of the form
\begin{equation} \label{eq:Appendix4_1}
\psi(x)=e^{-ip.x}u,
\end{equation}
for some $p\in{\bf R}^4,u\in{\bf C}^4$. If $u=0$ the the Dirac equation is trivially satisfied, so assume that $u\neq0$. Now if $\psi$ is of this form then
\begin{equation}
(i{\slas\partial}-m)\psi=0\Leftrightarrow({\slas p}-m)u=0.
\end{equation}
If this is the case then
\begin{equation}
0=({\slas p}+m)({\slas p}-m)u=(p^2-m^2)u.
\end{equation}
Therefore we must have that $p^2=m^2$, i.e. that $p\in H_{\pm m}$.
Thus we are seeking $p\in H_{\pm m}, u\in{\bf C}^4\backslash\{0\}$ such that $({\slas p}-m)u=0$, i.e. $u\in\mbox{Ker}({\slas p}-m)$. 

Let $p\in H_{\pm m}$. Choose $\Lambda\in O(1,3)^{+\uparrow},\kappa\in K$ such that 
\begin{equation}
\Lambda p=(\pm m,{\vct 0})^{T}, \Lambda=\Lambda(\kappa),
\end{equation}
where $\Lambda(\kappa)$ is the Lorentz transformation corresponding to $\kappa\in K$. (see (Mashford, 2017a)). Then
\begin{eqnarray}
\mbox{Ker}({\slas p}-m) & = & \kappa^{-1}\mbox{Ker}(\kappa({\slas p}-m)\kappa^{-1}) \nonumber \\
    & = & \kappa^{-1}\mbox{Ker}(\Sigma(\Lambda p)-m) \nonumber \\
    & = & \kappa^{-1}\mbox{Ker}(\Sigma((\pm m,{\vct 0})^{T})-m), \nonumber
\end{eqnarray}
where $\Sigma$ denotes the map $p\mapsto{\slas p}$. 

We will use the Dirac representation for the gamma matrices in which 
\begin{equation}
\gamma^0=\left(\begin{array}{cc}
1 & 0 \\
0 & -1
\end{array}\right).
\end{equation}
With respect to the metric $g=\gamma^0$ the vectors $\{e_{\alpha}\}_{\alpha=0}^3$ form an orthonormal basis where
\begin{equation}
(e_{\alpha})_{\beta}=\delta_{\alpha\beta},
\end{equation}
i.e.
\begin{equation}
\overline{e}_{\alpha}e_{\beta}=e_{\alpha}^{\dagger}\gamma^0e_{\beta}=\gamma^0_{\alpha\beta},\forall\alpha,\beta\in\{0,1,2,3\}.
\end{equation}
Now
\begin{equation}
\Sigma((\pm m,{\vct 0})^{T})-m=\left(\begin{array}{cc}
\pm m-m & 0 \\
0 & \mp m-m
\end{array}\right),
\end{equation}
Therefore, if $u=(u_1,u_2)^{T}$ then
\begin{equation}
u\in\mbox{Ker}(\Sigma((\pm m,{\vct 0})^{T})-m) \Leftrightarrow \left(\begin{array}{cc}
\pm m-m & 0 \\
0 & \mp m-m
\end{array}\right)\left(\begin{array}{l}
u_1 \\
u_2
\end{array}\right)=0 \nonumber
\end{equation}
In the positive energy case, i.e. when $p\in H_m$ this is equivalent to 
\begin{equation}
u_1=\mbox{ arbitrary}, u_2=0.
\end{equation}
Hence Dim$(\mbox{Ker}({\slas p}-m))=2$. In other words fermions have 2 polarization states. A basis for $\mbox{Ker}({\slas p}-m)$ is 
\begin{equation}
u_0=\kappa^{-1}e_0, u_1=\kappa^{-1}e_1,
\end{equation}
and we may describe $u_0,u_1$ as being Dirac spinors associated with $p\in H_m$ ($u_0,u_1$ are not unique because the choice of $\kappa$ is not unique).

Similarly, in the negative energy case, i.e. when $p\in H_{-m}$ a basis for  $\mbox{Ker}({\slas p}-m)$ is 
\begin{equation}
v_0=\kappa^{-1}e_2, v_1=\kappa^{-1}e_3.
\end{equation}

Now let $v\in{\bf C}^4$. Then clearly $({\slas p}+m)v\in\mbox{Ker}({\slas p}-m)$. Therefore the space $<({\slas p}+m)e_{\alpha},\alpha=0,1,2,3>$ is a subspace of Ker$({\slas p}-m)$. We will show that in fact it is equal to Ker$({\slas p}-m)$. We have
\begin{eqnarray}
({\slas p}+m) & = & \kappa^{-1}\kappa({\slas p}+m)\kappa^{-1}\kappa \nonumber \\
    & = & \kappa^{-1}(\Sigma((\pm m,{\vct 0})^{T}+m)\kappa  \nonumber \\
    & = & \kappa^{-1}\left(\begin{array}{ll}
\pm m+m & 0 \\
0 & \mp m+m
\end{array}\right)\kappa \nonumber \\
 & = & \kappa^{-1}\left(\begin{array}{cccc}
\pm m+m & 0 & 0 & 0 \\
0 & \pm m+m & 0 & 0 \\
0 & 0 & \mp m+m & 0 \\
0 & 0 & 0 & \mp m+m
\end{array}\right)\kappa. \nonumber
\end{eqnarray}
Thus, in the positive energy case,
\begin{equation}
({\slas p}+m)=2m\kappa^{-1}(e_0,e_1,0,0)\kappa,
\end{equation}
and in the negative energy case
\begin{equation}
({\slas p}+m)=2m\kappa^{-1}(0,0,e_2,e_3)\kappa.
\end{equation}
Therefore
\begin{equation}
\frac{{\slas p}+m}{2m}=(u_0,u_1,0,0)\kappa,
\end{equation}
(positive energy) and
\begin{equation}
\frac{{\slas p}+m}{2m}=(0,0,v_0,v_1)\kappa,
\end{equation}
(negative energy).

Let $w_{\alpha}=\kappa^{-1}e_{\alpha}, \alpha=0,1,2,3$. $\{w_{\alpha}\}_{\alpha=0}^3$ forms an orthonormal basis for ${\bf C}^4$ with respect to the metric $g=\gamma^0$. Then
\begin{equation}
\frac{{\slas p}+m}{2m}w_{\alpha}=u_{\alpha},\mbox{ for }\alpha=0,1,
\end{equation}
(positive energy) and
\begin{equation}
\frac{{\slas p}+m}{2m}w_{\alpha+2}=v_{\alpha},\mbox{ for }\alpha=0,1,
\end{equation}
(negative energy).

Since $\{(2m)^{-1}({\slas p}+m)w_{\alpha},\alpha=0,1\}$ is a basis for Ker$({\slas p}-m)$ it follows that $\{(2m)^{-1}({\slas p}+m)e_{\alpha},\alpha=0,1,2,3\}$ spans Ker$({\slas p}-m)$.

It is straightforward to show that the Dirac spinors that we have constructed satisfy the usual normalization properties (Itzykson and Zuber, 1980, p. 696).

\subsection*{Dirac bilinears in the non-relativistic approximation}

In the non-relativistic approximation we have
\begin{equation}
\frac{p^0}{m}\approx1,\frac{p^j}{m}\approx0,\mbox{ for } j=1,2,3.
\end{equation}
Therefore
\begin{equation}
\kappa\approx I.
\end{equation}
Therefore we can take, in the positive energy case,
\begin{equation}
(u_0,u_1,0,0)=\frac{{\slas p}+m}{2m}\kappa^{-1}=\frac{1}{2}(\Sigma((1,{\vct 0})^{T})+1)=\left(\begin{array}{cccc}
1 & 0 & 0 & 0 \\
0 & 1 & 0 & 0 \\
0 & 0 & 0 & 0 \\
0 & 0 & 0 & 0 
\end{array}\right).
\end{equation}
Thus 
\begin{equation}
u_0(p)=e_0,u_1(p)=e_1,\forall p\in H_m.
\end{equation}
Therefore
\begin{equation}
\overline{u}_{\alpha}(p^{\prime})\gamma^0u_{\beta}(p)=u^{\dagger}_{\alpha}(p^{\prime})\gamma^0\gamma^0u_{\beta}(p)=e_{\alpha}^{\dagger}e_{\beta}=\delta_{\alpha\beta},\forall\alpha,\beta\in\{0,1\},p,p^{\prime}\in H_m.
\end{equation}
Also
\begin{eqnarray}
\overline{u}_{\alpha}(p^{\prime})a_{j}\gamma^{j}u_{\beta}(p) & = & u^{\dagger}_{\alpha}(p^{\prime})\gamma^{0}a_{j}\gamma^{j}u_{\beta}(p) \nonumber \\
    & = & e_{\alpha}^{\dagger}\left(\begin{array}{cccc}
1 & 0 & 0 & 0 \\
0 & 1 & 0 & 0 \\
0 & 0 & -1 & 0 \\
0 & 0 & 0 & -1
\end{array}\right)\left(\begin{array}{cccc}
0 & 0 & a_3 & a_1-ia_2 \\
0 & 0 & a_1+ia_2 & -a_3 \\
-a_3 & -a_1+ia_2 & 0 & 0 \\
-a_1-ia_2 & a_3 & 0 & 0
\end{array}\right)e_{\beta} \nonumber \\
    & = 0, \nonumber
\end{eqnarray}
for all $\alpha,\beta\in\{0,1\},a\in{\bf R}^4,p,p^{\prime}\in H_m$. Therefore
\begin{equation}
\overline{u}_{\alpha}(p^{\prime})\gamma^{j}u_{\beta}(p)=0,\forall\alpha,\beta\in\{0,1\},j\in\{1,2,3\},p,p^{\prime}\in H_m.
\end{equation}

\section*{Appendix 4: Rigorous justification of Argument 1}

The following theorem establishes that Argument 1 is justified.
\begin{theorem} \label{theorem:Argument_1}
Let $g(a,b,\epsilon)$ be defined by $g(a,b,\epsilon)=\mu(\Gamma(a,b,\epsilon))\mbox{ for }a,b\in{\bf R},a<b,\epsilon>0$, where $\mu=\Omega_m*\Omega_m$. Then the following formal argument (Argument 1)
\begin{eqnarray}
g(a,b,\epsilon) & = & \mu(\Gamma(a,b,\epsilon)) \nonumber \\
    & = & \int\chi_{\Gamma(a,b,\epsilon)}(p+q)\,\Omega_m(dp)\,\Omega_m(dq) \nonumber \\
    & \approx & \int\chi_{(a,b)\times B_{\epsilon}(0)}(p+q)\,\Omega_m(dp)\,\Omega_m(dq) \nonumber \\
    & = & \int\chi_{(a,b)}(\omega_m({\vct p})+\omega_m({\vct q}))\chi_{B_{\epsilon}(0)}({\vct p}+{\vct q})\,\frac{d{\vct p}}{\omega_m({\vct p})}\frac{d{\vct q}}{\omega_m({\vct q})} \nonumber \\
    & = & \int\chi_{(a,b)}(\omega_m({\vct p})+\omega_m({\vct q}))\chi_{B_{\epsilon}(0)-{\vct q}}({\vct p})\,\frac{d{\vct p}}{\omega_m({\vct p})}\frac{d{\vct q}}{\omega_m({\vct q})} \nonumber \\
    & \approx & \int\chi_{(a,b)}(2\omega_m({\vct q}))\frac{\frac{4}{3}\pi\epsilon^3}{\omega_m({\vct q})^2}\,d{\vct q}, \nonumber
\end{eqnarray}
is justified in the sense that
\begin{equation}
\lim_{\epsilon\rightarrow0}\epsilon^{-3}g(a,b,\epsilon)=\frac{4}{3}\pi\int\chi_{(a,b)}(2\omega_m({\vct q}))\frac{1}{\omega_m({\vct q})^2}\,d{\vct q}.
\end{equation}
\end{theorem}
{\bf Proof}\\
There are 2 $\approx$ signs that we have to consider. The~first is in line 3 and arises because we are approximating the hyperbolic cylinder of radius $\epsilon$ between $a$ and $b$ with an ordinary cylinder of radius $\epsilon$. We will show that the error is of order greater than $\epsilon^3$. Let $\Gamma=\Gamma(a,b,\epsilon)$ be the aforementioned hyperbolic cylinder. Then
\begin{equation}
\Gamma=\bigcup_{m^{\prime}\in(a,b)}S(m^{\prime},\epsilon).
\end{equation}
Let
\begin{eqnarray}
\Gamma^{\prime} & = & \bigcup_{m^{\prime}\in(a,b)}\{m^{\prime}\}\times B_{\epsilon}({\vct 0}) =(a,b)\times B_{\epsilon}({\vct 0})\nonumber \\
\Gamma^{\prime-} & = & \bigcup_{m^{\prime}\in(a,a^{+})}\{(m^{\prime},{\vct p}):{\vct p}^2>m^{\prime2}-a^2\} \nonumber \\
    &  \subset & \bigcup_{m^{\prime}\in(a,a^{+})}(\{m^{\prime}\}\times B_{\epsilon}({\vct 0}))=(a,a^{+})\times B_{\epsilon}({\vct 0}) \nonumber \\
\Gamma^{\prime+} & = & \bigcup_{m^{\prime}\in(b,b^{+})}\{(m^{\prime},{\vct p}):{\vct p}^2>m^{\prime2}-b^2\} \nonumber \\
    &  \subset & \bigcup_{m^{\prime}\in(b,b^{+})}(\{m^{\prime}\}\times B_{\epsilon}({\vct 0}))=(b,b^{+})\times B_{\epsilon}({\vct 0}), \nonumber
\end{eqnarray}
in which
\begin{equation}
a^{+}=(a^2+\epsilon^2)^{\frac{1}{2}}, b^{+}=(b^2+\epsilon^2)^{\frac{1}{2}}.
\end{equation}
Then $\Gamma$ differs from $(\Gamma^{\prime}\sim\Gamma^{\prime-})\cup\Gamma^{\prime+}$ on a set of measure zero,

It is straightforward to show that if $\Gamma_1,\Gamma_2\in{\mathcal B}_0({\bf R}^4),\Gamma_1\cap\Gamma_2=\emptyset$ then
\begin{eqnarray}
\int\chi_{\Gamma_1\cup\Gamma_2}(p+q)\,\Omega_m(dp)\,\Omega_m(dq) & = & \int\chi_{\Gamma_1}(p+q)\,\Omega_m(dp)\,\Omega_m(dq)+ \nonumber \\ 
    &  & \int\chi_{\Gamma_2}(p+q)\,\Omega_m(dp)\,\Omega_m(dq). \nonumber
\end{eqnarray}
Therefore
\begin{eqnarray}
& & |\int\chi_{\Gamma}(p+q)\,\Omega_m(dp)\,\Omega_m(dq)-\int\chi_{\Gamma^{\prime}}(p+q)\,\Omega_m(dp)\,\Omega_m(dq)| \leq \nonumber \\
& & \int\chi_{\Gamma^{\prime-}}(p+q)\,\Omega_m(dp)\,\Omega_m(dq)+\int\chi_{\Gamma^{\prime+}}(p+q)\,\Omega_m(dp)\,\Omega_m(dq). \nonumber
\end{eqnarray} 
We will show that
\begin{equation}
\lim_{\epsilon\rightarrow0}(\epsilon^{-3}\int\chi_{\Gamma^{\prime\pm}}(p+q)\,\Omega_m(dp)\Omega_m(dq))=0.
\end{equation}
It suffices to consider the $-$ case.
We have
\begin{equation}
\begin{aligned} \label{eq:first_approx1} 
\int\chi_{\Gamma^{\prime-}}(p+q)\,\Omega_m(dp)\Omega_m(dq)  \leq &~\int\chi_{(a,a^{+})\times B_{\epsilon}(0)}(p+q)\,\Omega_m(dp)\,\Omega_m(dq)  \\
     = &~\int\chi_{(a,a^{+})}(\omega_m({\vct p})+\omega_m({\vct q}))\chi_{B_{\epsilon}(0)-{\vct q}}({\vct p})\,\frac{d{\vct p}}{\omega_m({\vct p})}  \\
     &~\frac{d{\vct q}}{\omega_m({\vct q})}. 
\end{aligned}
\end{equation}
We will come back to this equation later but will now return to the general argument Argument 1 and consider the second and final $\approx$. This $\approx$ arises because we are approximating ${\vct p}$ by $-{\vct q}$ since ${\vct p}$ ranges over a ball of radius $\epsilon$ centred on $-{\vct q}$.

Suppose that ${\vct p}$ and ${\vct q}$ are such that $\chi_{B_{\epsilon}(0)-{\vct q}}({\vct p})=1$. Then $|{\vct p}+{\vct q}|<\epsilon$. Thus $||{\vct p}|-|{\vct q}||<\epsilon$. Hence
\begin{eqnarray}
|\omega_m({\vct p})-\omega_m({\vct q})| & = & |({\vct p}^2+m^2)^{\frac{1}{2}}-({\vct q}^2+m^2)^{\frac{1}{2}})| \nonumber \\
    & =  & \left|\frac{{\vct p}^2-{\vct q}^2}{({\vct p}^2+m^2)^{\frac{1}{2}}+({\vct q}^2+m^2)^{\frac{1}{2}})}\right| \nonumber \\
    & \leq & \frac{|{\vct p}^2-{\vct q}^2|}{2m} \nonumber \\
    & = & \frac{||{\vct p}|-|{\vct q}||(|{\vct p}|+|{\vct q}|)}{2m} \nonumber \\
    & < & \frac{\epsilon}{2m}(|{\vct p}|+|{\vct q}|). \nonumber
\end{eqnarray}
We have $|{\vct p}|\in(|{\vct q}|-\epsilon,|{\vct q}|+\epsilon)$. Therefore $|{\vct p}|+|{\vct q}|<2|{\vct q}|+\epsilon$. Thus 
\[ |\omega_m({\vct p})-\omega_m({\vct q})| < \frac{\epsilon}{2m}(2|{\vct q}|+\epsilon). \]
Therefore
\begin{equation}
\begin{aligned}
\omega_m({\vct p})+\omega_m({\vct q})  = &~\omega_m({\vct p})-\omega_m({\vct q})+\omega_m({\vct q})+\omega_m({\vct q})  \\
     \leq &~|\omega_m({\vct p})-\omega_m({\vct q})|+2\omega_m({\vct q})  \\
     < &~2\omega_m({\vct q})+ \frac{\epsilon}{2m}(2|{\vct q}|+\epsilon). \label{equation:omega_sum}
\end{aligned}
\end{equation}
Now let
\begin{eqnarray}
I(\epsilon) & = & \int\chi_{(a,b)}(\omega_m({\vct p})+\omega_m({\vct q}))\chi_{B_{\epsilon}({\vct0})-{\vct q}}({\vct p})\,\frac{d{\vct p}}{\omega_m({\vct p})}\frac{d{\vct q}}{\omega_m({\vct q})} \nonumber \\
J(\epsilon) & = & \int\chi_{(a,b)}(2\omega_m({\vct q}))\chi_{B_{\epsilon}({\vct0})-{\vct q}}({\vct p})\,\frac{d{\vct p}}{\omega_m({\vct p})}\frac{d{\vct q}}{\omega_m({\vct q})} \nonumber \\  
K(\epsilon) & = &  \int\chi_{(a,b)}(2\omega_m({\vct q}))\frac{\frac{4}{3}\pi\epsilon^3}{\omega_m({\vct q})^2}\,d{\vct q}. \nonumber
\end{eqnarray}
We will show that
\begin{equation} \label{eq:Appendix6_limits1}
\lim_{\epsilon\rightarrow0}\epsilon^{-3}(I(\epsilon)-J(\epsilon))=0,
\mbox{ and }
\lim_{\epsilon\rightarrow0}\epsilon^{-3}(J(\epsilon)-K(\epsilon))=0.
\end{equation}
Concerning the first limit we note that $\chi_{(a,b)}(\omega_m({\vct p})+\omega_m({\vct q}))$ differs from $\chi_{(a,b)}(2\omega_m({\vct q}))$ if and only~if 
\begin{enumerate}
\item $\omega_m({\vct p})+\omega_m({\vct q})\in(a,b)$ but $2\omega_m({\vct q})\leq a$ or 
\item $\omega_m({\vct p})+\omega_m({\vct q})\in(a,b)$ but $2\omega_m({\vct q})\geq b$ or
\item $2\omega_m({\vct q})\in(a,b)$ but $\omega_m({\vct p})+\omega_m({\vct q})\leq a$ or
\item $2\omega_m({\vct q})\in(a,b)$ but $\omega_m({\vct p})+\omega_m({\vct q})\geq b$.
\end{enumerate}
Thus
\begin{equation}
|I(\epsilon)-J(\epsilon)|=I_1(\epsilon)+I_2(\epsilon)+I_3(\epsilon)+I_4(\epsilon),
\end{equation}
where
\begin{equation}
\begin{aligned}
I_1(\epsilon)  = &~\int\chi_{(a,b)}(\omega_m({\vct p})+\omega_m({\vct q}))\chi_{(-\infty,a]}(2\omega_m({\vct q}))\chi_{B_{\epsilon}({\vct0})-{\vct q}}({\vct p})\,  \\
     &~\frac{d{\vct p}}{\omega_m({\vct p})}\frac{d{\vct q}}{\omega_m({\vct q})},  \\
\end{aligned}
\end{equation}
and $I_2,I_3,I_4$ are defined similarly.
We will show that
\begin{equation}
\lim_{\epsilon\rightarrow0}\epsilon^{-3}I_1(\epsilon)=0.
\end{equation}
$I_2,I_3$ and $I_4$ can be dealt with~similarly. 

Using Equation~(\ref{equation:omega_sum})
\[ \omega_m({\vct p})+\omega_m({\vct q})\in(a,b)\mbox{ and }2\omega_m({\vct q})\leq a\Rightarrow a-\frac{\epsilon}{2m}(2|{\vct q}|+\epsilon)<2\omega_m({\vct q})\leq a. \]
Therefore
{\begin{eqnarray}
I_1(\epsilon) & \leq & \int\chi_{(a-(2|{\vct q}|+\epsilon)\epsilon/(2m),a]}(2\omega_m({\vct q}))\chi_{B_{\epsilon}({\vct0})-{\vct q}}({\vct p})\frac{1}{m^2}\,d{\vct p}\,d{\vct q}\nonumber \\
& = & \frac{4}{3}\pi\epsilon^3\int\chi_{(a-(2|{\vct q}|+\epsilon)\epsilon/(2m),a]}(2\omega_m({\vct q}))\frac{1}{m^2}\,d{\vct q}.\nonumber
\end{eqnarray}
Hence
\begin{eqnarray}
\epsilon^{-3}I_1(\epsilon) & \leq & \frac{4}{3}\pi\int\chi_{(a-(2|{\vct q}|+\epsilon)\epsilon/(2m),a]}(2\omega_m({\vct q}))\frac{1}{m^2}\,d{\vct q}.
\end{eqnarray}
The integrand is integrable for all $\epsilon>0$, vanishes outside the compact set
\[ C = \{{\vct q}\in{\bf R}^3:2\omega_m({\vct q})\leq a\}, \]
is dominated by the integrable function
\[ g({\vct q})=\frac{1}{m^2}\chi_{[0,a]}(2\omega_m({\vct q})), \]
and converges pointwise to $0$ everywhere on ${\bf R}^3$ as $\epsilon\rightarrow0$ except on the set $\partial C=\{{\vct q}\in{\bf R}^3:2\omega_m({\vct q})=a\}$ which is a set of measure $0$. Therefore by the dominated convergence theorem
\begin{equation}
\lim_{\epsilon\rightarrow0}\epsilon^{-3}I_1(\epsilon)=0,
\end{equation}
as~required.

Now regarding the second limit in Equation~(\ref{eq:Appendix6_limits1}) consider the function $f:[0,\infty)\rightarrow(0,m^{-1}]$ defined by
\begin{equation}
f(p)=(m^2+p^2)^{-\frac{1}{2}}.
\end{equation}
$f$ is analytic. Therefore by Taylor's theorem for all $q,p\ge0$
\begin{equation}
f(p)=f(q)+f^{\prime}(q)(p-q)+\frac{1}{2}f^{\prime\prime}(\xi)(p-q)^2,
\end{equation}
for some $\xi$ between $q$ and $p$. Now
\begin{align*}
f^{\prime}(p)&=-p(m^2+p^2)^{-\frac{3}{2}}\\
f^{\prime\prime}(p)&=(m^2+p^2)^{-\frac{5}{2}}(2p^2-m^2).
\end{align*}
Therefore
\begin{align*}
|f^{\prime\prime}(\xi)|&=(m^2+\xi^2)^{-\frac{5}{2}}|2\xi^2-m^2|\\
&\leq m^{-5}(2(q+\epsilon)^2+m^2),
\end{align*}
as long as $|p-q|<\epsilon$. Thus
\begin{align*}
|f(p)-f(q)|&=|f^{\prime}(q)(p-q)+\frac{1}{2}f^{\prime\prime}(\xi)(p-q)^2|\\
&<q(m^2+q^2)^{-\frac{3}{2}}\epsilon+\frac{1}{2}m^{-5}(2(q+\epsilon)^2+m^2)\epsilon^2\\
&<m^{-1}\epsilon+\frac{1}{2}m^{-5}(2(q+\epsilon)^2+m^2)\epsilon^2,
\end{align*}
as long as $|p-q|<\epsilon$. Hence
\begin{align*}
|J(\epsilon)-K(\epsilon)|&=|\int\chi_{(a,b)}(2\omega_m({\vct q}))(\int\chi_{B_{\epsilon}({\vct0})-{\vct q}}({\vct p})(\frac{1}{\omega_m({\vct p})}-\frac{1}{\omega_m({\vct q})})\,d{\vct p})\frac{d{\vct q}}{\omega_m({\vct q})}|\\
&\leq\int\chi_{(a,b)}(2\omega_m({\vct q}))(\int\chi_{B_{\epsilon}({\vct0})-{\vct q}}({\vct p})(|f(|{\vct p}|)-f(|{\vct q}|)|)\,d{\vct p})\frac{d{\vct q}}{\omega_m({\vct q})}\\
&\leq\int\chi_{(a,b)}(2\omega_m({\vct q}))\int\chi_{B_{\epsilon}({\vct0})-{\vct q}}({\vct p})(m^{-1}\epsilon+\frac{1}{2}m^{-5}(2(|{\vct q}|+\epsilon)^2+m^2)\epsilon^2)\\
&\,d{\vct p}\frac{d{\vct q}}{\omega_m({\vct q})}\\
&=\frac{4}{3}\pi\epsilon^3\int\chi_{(a,b)}(2\omega_m({\vct q}))(m^{-1}\epsilon+\frac{1}{2}m^{-5}(2(|{\vct q}|+\epsilon)^2+m^2)\epsilon^2)\frac{d{\vct q}}{\omega_m({\vct q})}.
\end{align*}
Therefore
\begin{align*}
\lim_{\epsilon\rightarrow0}\epsilon^{-3}|J(\epsilon)-K(\epsilon)|&=\lim_{\epsilon\rightarrow0}\frac{4}{3}\pi\int\chi_(a,b)(2\omega_m({\vct q}))(m^{-1}\epsilon+\frac{1}{2}m^{-5}(2(|{\vct q}|+\epsilon)^2+m^2)\epsilon^2)\\
&\frac{d{\vct q}}{\omega_m({\vct q})}\\
&=0,
\end{align*}
as required. We have therefore dealt with the second $\approx$ in Argument~1. 

To finish dealing with the first $\approx$ suppose that $\epsilon_1>0$ is given. Choose $c\in(a,b)$ such that
\begin{equation}
\frac{16}{3}\pi^2(Z(c)-Z(a))<\frac{\epsilon_1}{2}.
\end{equation}
Now choose $\delta_1>0$ such that if $0<\epsilon<\delta_1$ then
\begin{eqnarray}
&  & |\epsilon^{-3}\int\chi_{(a,c)}(\omega_m({\vct p})+\omega_m({\vct q}))\chi_{B_{\epsilon}({\vct 0})-{\vct q}}({\vct p})\frac{d{\vct p}}{\omega_m({\vct p})}\frac{d{\vct q}}{\omega_m({\vct q})}- \nonumber \\
  &  & \frac{4}{3}\pi\int\chi_{(a,c)}(2\omega_m({\vct q}))\frac{d{\vct q}}{\omega_m({\vct q})^2}|<\frac{\epsilon_1}{2}. \nonumber
\end{eqnarray}
(We can do this because of the validity of the second $\approx$.)
Choose $\delta_2>0$ such that if $0<\epsilon<\delta_2$ then $a^{+}=a^{+}(\epsilon)<c$. Let $\delta=\mbox{min}(\delta_1,\delta_2)$. 

Then if $\epsilon<\delta$ then
\begin{eqnarray}
  &  & |\epsilon^{-3}\int\chi_{(a,a^{+})}(\omega_m({\vct p})+\omega_m({\vct q}))\chi_{B_{\epsilon}({\vct 0})-{\vct q}}({\vct p})\frac{d{\vct p}}{\omega_m({\vct p})}\frac{d{\vct q}}{\omega_m({\vct q})}| \nonumber \\
  & \leq & |\epsilon^{-3}\int\chi_{(a,c)}(\omega_m({\vct p})+\omega_m({\vct q}))\chi_{B_{\epsilon}({\vct 0})-{\vct q}}({\vct p})\frac{d{\vct p}}{\omega_m({\vct p})}\frac{d{\vct q}}{\omega_m({\vct q})}| \nonumber \\
  & \leq &  |\epsilon^{-3}\int\chi_{(a,c)}(\omega_m({\vct p})+\omega_m({\vct q}))\chi_{B_{\epsilon}({\vct 0})-{\vct q}}({\vct p})\frac{d{\vct p}}{\omega_m({\vct p})}\frac{d{\vct q}}{\omega_m({\vct q})}- \nonumber \\
  &  & \frac{4}{3}\pi\int\chi_{(a,c)}(2\omega_m({\vct q}))\frac{d{\vct q}}{\omega_m({\vct q})^2}|+|\frac{4}{3}\pi\int\chi_{(a,c)}(2\omega_m({\vct q}))\frac{d{\vct q}}{\omega_m({\vct q})^2}| \nonumber \\
  & < & \frac{\epsilon_1}{2}+\frac{4}{3}\pi\int\chi_{(a,c)}(2\omega_m({\vct q}))\frac{d{\vct q}}{\omega_m({\vct q})^2} \nonumber \\
  & = & \frac{\epsilon_1}{2}+\frac{4}{3}\pi\int_{Z(a)}^{Z(c)}\frac{4\pi r^2}{m^2+r^2}\,dr \nonumber \\
  & \leq & \frac{\epsilon_1}{2}+\frac{16}{3}\pi^2(Z(c)-Z(a)) \nonumber \\
  & < & \frac{\epsilon_1}{2}+\frac{\epsilon_1}{2} \nonumber \\
  & = & \epsilon_1. \nonumber
\end{eqnarray}
Thus
\begin{equation}
\lim_{\epsilon\rightarrow0}\epsilon^{-3}\int\chi_{(a,a^{+})}(\omega_m({\vct p})+\omega_m({\vct q}))\chi_{B_{\epsilon}({\vct 0})-{\vct q}}({\vct p})\frac{d{\vct p}}{\omega_m({\vct p})}\frac{d{\vct q}}{\omega_m({\vct q})}=0,
\end{equation}
thereby completing the proof of the validity of the first $\approx$ and therefore the validity of Argument 1. $\Box$

\end{document}